\documentclass{article}
\usepackage{psfig}
  
\begin{document}
  
  \title{A simplified model of the formation of structures in the dark matter, and a background of very long gravitational waves }
  \author{G.S. Bisnovatyi-Kogan
  \thanks{Space Research Institute, Moscow, Russia, Profsoyuznaya
  84/32, Moscow 117810, Russia;  gkogan@mx.iki.rssi.ru; and Theoretical Astrophysics Center, Copenhagen, Denmark}}
  \date{}
  \maketitle
  \begin{abstract}
  Collapse of the rotating spheroid is approximated by a system of ordinary differential equations describing its dynamics. The gravitational potential is approximated by the one of the iniform Maclaurin spheroid. 
 Developement of gravitational instability and collapse in the dark matter medium do not lead to any shock formation or radiation, but is characterized by non-collisional relaxation, which is accompanied by the mass and angular momentum losses. Phenomenological account of these processes is done in this model. Formation of the equilibrium configuration dynamics of collapse is investigated for several parameters, characterizing the configuration. A very long gravitational wave emission during the collapse is estimated, and their possible connection with the observed gravitational lenses is discussed.   
\end{abstract}

\section{Introduction}

A study of of a formation of dark matter objects in the universe is based on 
N-body simulations, which are very time consuming. In this situation a simplified approach may become useful, because it permits to investigate rapidly by analytical or by simple numerical calculations many different variants , and obtain crudly some new principally important features of the problem which could be lost or not visible during a long numerical  work.

The modern theory of a large scale structure is based on the ideas of Zeldovich (1970) about a formation of strongly non-spherical structures during non-linear stages of a development of the gravitational instability, known as "Zeldovich's pancakes". The numerical simulations started by Doroshkevich et al. (1980) had been performed subsequently by many groups, revealing complicated structures of the new born objects (see e.g. Doroshkevich et al., 1999).

Here we derive and solve equations for a simple model of a dynamical behaviour of a compressible rotating spheroid in which a motion along both axies takes place in the common gravitational field of a uniform spheroid, and under the action of the isotropic pressure (taken into account in the approximate non-differential way), and otherwise independently on each other. 
Due to non-interacting particles of the dark matter a strong compression does not lead to a formation of the shock wave and additional energy losses due to radiation at increasing of a temperature.. All losses are connected with a runaway particles, and relaxation consists only in the phase mixing leading effectively to a transformation of the kinetic energy of the ordered motion (collapse) into the kinetic energy of the chaotic motion, and to creation an effective pressure and thermal energy. In presence of the chaotic motion the pancake formed during the collapse has a finite minimal thickness, because particles cross the equatorial plane non-simultaneously, forming therefore the effective pressure, which "stops" the contraction. These effects may be described approximately in the hydrodynamical approach, in which there are no shock waves and the main transport process is an effective bulk viscosity, leading to relaxation and to a damping of the ordered motion.
This relaxation is based on the idea of a "violent relaxation" of Lynden-Bell (1967). The system of the ordinary differential equations is derived, which describes the dynamical behaviour of the compressible spheroid, where the relaxation and losses of the energy, mass, and angular momentum are taken into accound phenomenologically. In absence of any dissipation these equations describe a self-consistent conservative system, where gravitational and thermal pressure forces are present. Namely, the equilibrium solution of these equation describes a Maclaurin spheriod, where the density is not prescribed but is found self-consistently, when the effective entropy is known. We consider here only non-relativistic dark matter with a relations between the density $\rho$, energy density $E$, pressure $P$, and specific entropy $S$ as

$$\rho E =\frac{3}{2}P, \quad P\sim \rho^{5/3}, \quad E\sim \rho^{2/3} \quad
{\rm at\,\,\, constant}\,\,\, S.
$$
Solution of simple ordinary equations reveals some interesting features of the collapse of non-interacting matter, which had not been noticed earlier. The collapse of a spherioid, leading to the formation of a pancake may be continued by the fornation of a transient oblate figure, strongly elongated along the axis of a symmetry with a ratio of the axis $c$ to the equatorial radius $a$ : $c/a \sim 5$. Formation of such transient figures, appearing only at low rotation, could be expected in central parts of the collapsing dark matter objects during initial stages of the collapse, when the influence of the outer non-uniform part of the cloud is still negligible.

We have found, that at a realistic rate of the relaxation rate with a characteristic time equal to 3 local Jeans times, the oscillations do not damp completely, and about 1/10 of the initial amplitude of a velocity survives after at least 10 oscillationl periods. That means that the most massive dark matter objects may be still in the oscillatory state. 
These oscillations, as well as variable gravitational fields of the collapsing dark matter objects 
 could be important for the the interpretation of the observed picture of the cosmic microvave background (CMB) fluctuations. Very long gravitational waves emitted mainly during the  
first stage of the pancake formation (see Thuan and Ostriker, 1974; Novikov, 1976) could be also important for the CMB fluctuations (Doroshkevich et al., 1977), as well as for the gravitational lensing of the distant objects. 

 \section{Properties of a Maclaurin spheroid}

Let us consider a spheroid with semi-axies $a=b>c$

\begin{equation}
\label{eq1}
\frac{x^2+y^2}{a^2}+\frac{z^2}{c^2}=1,
\end{equation}
and rotating uniformly with an angular velocity $\Omega$.
Let us approximate the density of matter $\rho$ in the spheroid as uniform. The mass $m$ and total angular momentum $M$ of the spheroid are connected with the angular velosity and semi-axies as (Landau, Lifshits, 1988)

\begin{equation}
\label{eq2}
m=\frac{4\pi}{3} a^2 c \rho, \quad M=\frac{2}{5} m a^2 \Omega.
\end{equation}
Introducing $k=c/a<1$, we may express the gravitational energy $U_g$ as

\begin{equation}
\label{eq3}
U_g=-\frac{3Gm^2}{5a}\frac{\arccos{k}}{\sqrt{1-k^2}},
\end{equation}
what gives $U=-3Gm^2/5a$ for a sphere.
The gravitational forces inside the uniform spheriod $F_x, \,\, F_y,\,\,F_z$ are expressed via the grvitational potential $\phi$ as ${\bf F}=-\nabla\phi$, and are defined by the relations

\begin{equation}
\label{eq4}
F_x=\frac{2\pi G\rho x}{1-k^2}\biggl[k^2-\frac{k\arccos{k}}{\sqrt{1-k^2}}\biggr],
\quad F_y=\frac{y}{x}F_x,
\end{equation}
$$
F_z=-\frac{4\pi G\rho z}{1-k^2}\biggl[1-\frac{k\arccos{k}}{\sqrt{1-k^2}}\biggr].
$$
The gravitational potential with normalization $\phi \rightarrow 0$ at infinity, inside the spheroid is expressed as

\begin{equation}
\label{eq5}
\phi=-\frac{1}{2}(xF_x+yF_y+zF_z)-2\pi G\rho a^2 \frac{k\arccos{k}}{\sqrt{1-k^2}}.
\end{equation}

During the process of relaxation a form of the spheroid changes between oblate and prolate ones. When $c>a=b$ the following relations are valid instead of (\ref{eq3})-(\ref{eq5})

\begin{equation}
\label{eq3p}
U_g=-\frac{3Gm^2}{5a}\frac{{\rm Arch}{k}}{\sqrt{k^2-1}},
\end{equation}

\begin{equation}
\label{eq4p}
F_x=-\frac{2\pi G\rho x}{k^2-1}\biggl[k^2-\frac{k{\rm Arch}{k}}{\sqrt{k^2-1}}\biggr],
\quad F_y=\frac{y}{x}F_x,
\end{equation}
$$
F_z=\frac{4\pi G\rho z}{k^2-1}\biggl[1-\frac{k{\rm Arch}{k}}{\sqrt{k^2-1}}\biggr].
$$
\begin{equation}
\label{eq5p}
\phi=-\frac{1}{2}(xF_x+yF_y+zF_z)-2\pi G\rho a^2 \frac{k{\rm Arch}{k}}{\sqrt{k^2-1}}.
\end{equation}
Here the positive branch of ${\rm Arch}{k}$ should be used.

\section{Equations of motion}

For a real Maclaurin equilibrium noncompressible spheroid the oblateness $k$ is connected with the angular velocity (or angular momentum) 

\begin{equation}
\label{eq6}
\frac{k(1+2k^2)}{(1-k^2)^{3/2}}\arccos{k}-\frac{3k^2}{1-k^2}=
\frac{\Omega^2}{2\pi G\rho}=\frac{25}{6G}\frac{M^2}{m^3}\frac{k}{a}. 
\end{equation}
We consider a case of non-collisional particles when the density varies in time,
and approximate the form and gravitational potential of the body by the form and gravitational potential of Maclaurin spheroid. Here pressure does not nesessary balance the garvitational and centrifugal forces. Consider a uniform spheroid
with a constant mass and angular momentum, and with a total "thermal" energy of non-relativistic dark matter particles $E_{th} \sim V^{-2/3} \sim (abc)^{-2/3}$ ($V$ is a volume of a spheroid, $a=b$). Equations of motion for this spheroid are written for the radial accelerations at the equator and at the pole for  
semiaxes $\ddot a, \,\ddot c$

\begin{equation}
\label{eq7}
\ddot{a}=\frac{25}{4}\frac{M^2}{m^2a^3}+\frac{3}{2}\frac{Gm}{a^2(1-k^2)}
\biggl[k-\frac{\arccos{k}}{\sqrt{1-k^2}}\biggr]+\frac{10 E_{th,in}}{3 am}
\left(\frac{a_{in}b_{in} c_{in}}{abc}\right)^{2/3},
\end{equation}

\begin{equation}
\label{eq8}
\ddot{c}=-3\frac{Gm}{a^2(1-k^2)}
\biggl[1-\frac{k\arccos{k}}{\sqrt{1-k^2)}}\biggr]+\frac{10 E_{th,in}}{3 cm}
\left(\frac{a_{in}b_{in} c_{in}}{abc}\right)^{2/3}. 
\end{equation}
Here $E_{th,in},\,\,\, a_{in}=b_{in},\,\,\,c_{in}$ are the initial values of corresponding parameters.
Equations (\ref{eq7}),(\ref{eq8}) describe the dynamics of the concervative system with a linear dependence of the velocity on the coordinates 

\begin{equation}
\label{eq9}
v_x=\frac{\dot a x}{a}, \quad, v_y=\frac{\dot b y}{b}, \quad v_z=\frac{\dot c z}{c},
\end{equation}
and Lagrange function

\begin{equation}
\label{eq10}
L=U_{kin}-U_{pot},\quad
U_{pot}=U_{rot}+U_g+E_{th},  
\end{equation}
$$U_{rot}= \frac{5}{4}\frac{M^2}{ma^2},\quad E_{th}=\frac{\cal E}{(abc)^{2/3}},\quad {\cal E}=E_{th,in}(a_{in}b_{in} c_{in})^{2/3}
$$
The function $U_g$ defined in (\ref{eq3}), and the rotational energy from (\ref{eq10}) should be used during the variation of $L$ in the form

 \begin{equation}
\label{eq10a}
U_g=-\frac{6Gm^2}{5(a+b)}\frac{\arccos{k'}}{\sqrt{1-k'^2}},\quad
k'=\frac{2c}{a+b}, \quad U_{rot}= \frac{5}{2}\frac{M^2}{m(a^2+b^2)}.
\end{equation}
The kinetic energy is equal to

\begin{equation}
\label{eq11}
U_{kin}=\frac{1}{2}\rho \int_V\biggl[\left(\frac{\dot a x}{a}\right)^2+
\left(\frac{\dot b y}{b}\right)^2+\left(\frac{\dot c z}{c}\right)^2\biggr]dx\,dy\,dz=
\frac{m}{10}({\dot a}^2+{\dot b}^2+{\dot c}^2).
\end{equation} 
Equations of motion (\ref{eq7}),(\ref{eq8}) are the Lagrange equations with the Lagrange function (\ref{eq10})\footnote{We consider here for simplicity that the relaxation is accompanied by the isotropization of the distribution function in the velociyu space. The violent relaxation has a non-collisional origin, and if the anizotropic instability has a lower increment the anizotropy in the velocity space may be preserved. In the limiting case of a pure anizotropic relaxation we have three (or two in our case) pressure and entropy functions, so that

$$E_{th}=m\left(\frac{{\cal E}_x}{a^2}+\frac{{\cal E}_y}{b^2}+\frac{{\cal E}_z}{c^2}\right), \quad P_x=\frac{2}{3}\rho\frac{{\cal E}_x}{a^2}, \quad
P_y=\frac{2}{3}\rho\frac{{\cal E}_y}{b^2},\quad P_z=\frac{2}{3}\rho\frac{{\cal E}_z}{c^2}. $$
We should have the terms $!0{\cal E}_x/a^3m$ and $!0{\cal E}_z/c^3m$ instead of the last terms in (\ref{eq7}),(\ref{eq8}), respectively. A corresponding changes should be done also in the equations determining increase of entropies, and different losses.}
Solutions of these equations describe either pure oscillations, or the total disruption, depending on the initial conditions. 
For the prolate spheroid the following relations are valid instead of  
(\ref{eq7}),(\ref{eq8}),(\ref{eq10a})

\begin{equation}
\label{eq7p}
\ddot{a}=\frac{25}{4}\frac{M^2}{m^2a^3}-\frac{3}{2}\frac{Gm}{a^2(k^2-1)}
\biggl[k-\frac{{\rm Arch}{k}}{\sqrt{k^2-1}}\biggr]+\frac{10 E_{th,in}}{3 am}
\left(\frac{a_{in}b_{in} c_{in}}{abc}\right)^{2/3},
\end{equation}

\begin{equation}
\label{eq8p}
\ddot{c}=3\frac{Gm}{a^2(k^2-1)}
\biggl[1-\frac{k{\rm Arch}{k}}{\sqrt{k^2-1)}}\biggr]+\frac{10 E_{th,in}}{3 cm}
\left(\frac{a_{in}b_{in} c_{in}}{abc}\right)^{2/3}, 
\end{equation}

 \begin{equation}
\label{eq10p}
U_g=-\frac{6Gm^2}{5(a+b)}\frac{{\rm Arch}{k'}}{\sqrt{k'^2-1}},\quad
k'=\frac{2c}{a+b}, \quad U_{rot}= \frac{5}{2}\frac{M^2}{m(a^2+b^2)}.
\end{equation}

\section{Equations of motion with dissipation}

In the reality there is a relaxation in the collisionless system, connected with a phase mixing which called "violent relaxation" (Lynden-Bell, 1967). This relaxation leads to a dissipation of the energy of the kinetic motion and increase of the chaotic (thermal) energy and pressure. As a result of this dissipation the kinetic motion will suffer from an effective drag force, which account is described phenomenologically by adding of the terms

\begin{equation}
\label{eq12}
-\frac{\dot{a}}{\tau_{rel}} \quad {\rm and} -\frac{\dot{c}}{\tau_{rel}} 
\end{equation}
in the right parts of equations (\ref{eq7}),(\ref{eq8}) respectively. We consider the nonrelativistic dark matter with the relation between the pressure $P$ and the thermal energy $E_{th}$ of the whole spheroid as 

\begin{equation}
\label{eq13}
E_{th}=\frac{3}{2}\frac{Pm}{\rho} .
\end{equation}
Dissipation (\ref{eq12}) leads to a heat production. Let us find a rate of a heat production leading to the growth of the entropy of the spheroid matter. Write equations of motion (\ref{eq7}),(\ref{eq8}) with account of 
(\ref{eq5}),(\ref{eq10}),(\ref{eq12}) separately for $a,\,\, b$ and $c$, allowing also the energy,  mass and angular momentum losses

\begin{equation}
\label{eq14}
\frac{1}{m}\frac{d}{dt}\left(\frac{m^2{\dot a}^2}{10}\right)= -{\dot a}\frac{\partial U_g}{\partial a}-{\dot a}\frac{\partial U_{rot}}{\partial a}-{\dot a}\frac{\partial E_{th}}{\partial a}-\frac{1}{5}m\frac{{\dot a}^2}{\tau_{rel}},
\end{equation}

\begin{equation}
\label{eq15}
\frac{1}{m}\frac{d}{dt}\left(\frac{m^2{\dot b}^2}{5}\right)= -{\dot b}\frac{\partial U_g}{\partial b}-{\dot b}\frac{\partial U_{rot}}{\partial b}-{\dot b}\frac{\partial E_{th}}{\partial b}-\frac{1}{5}m\frac{{\dot b}^2}{\tau_{rel}},
\end{equation}

\begin{equation}
\label{eq16}
\frac{1}{m}\frac{d}{dt}\left(\frac{m^2{\dot c}^2}{5}\right)= -{\dot c}\frac{\partial U_g}{\partial c}-{\dot c}\frac{\partial U_{rot}}{\partial c}-{\dot c}\frac{\partial E_{th}}{\partial c}-\frac{1}{5}m\frac{{\dot c}^2}{\tau_{rel}},
\end{equation}

\begin{equation}
\label{eq17}
\tau_{rel}=2\pi\alpha_{rel}\sqrt{\frac{abc}{3Gm}}.
\end{equation}
Note that in (\ref{eq14})-(\ref{eq16}) the gravitational energy should be taken in the form (\ref{eq10a}).
We have scaled here the relaxation time $\tau_{rel}$ by the Jeans characteristic time

\begin{equation}
\label{eq18}
\tau_J=\frac{2\pi}{\omega_J}=\frac{2\pi}{\sqrt{4\pi Gm}}=
 2\pi\sqrt{\frac{abc}{3Gm}},
\end{equation}
with a constant value of $\alpha_{rel}$. Collecting together (\ref{eq14})-(\ref{eq16}) we find the energy balance relation in the form

\begin{equation}
\label{eq19}
\frac{dU_{tot}}{dt}=\frac{\partial U_g}{\partial m}\frac{dm}{dt}+
\frac{\partial U_{rot}}{\partial m}\frac{dm}{dt}+\frac{\partial U_{rot}}{\partial M}\frac{dM}{dt}+
\frac{\partial E_{th}}{\partial {\cal E}}\frac{d{\cal E}}{dt}-2\frac{U_{kin}}{\tau_{rel}}-\frac{U_{kin}}{m}\frac{dm}{dt},
\end{equation}
$$U_{tot}=U_{kin}+U_g+U_{rot}+E_{th}.
$$
The process of relaxation is accompanied also by the energy, mass, and angular momentum losses from the system. We suggest, that these losses take place only in non-stationary phases, so these rates are proportional to the kinetic energy $U_{kin}$,  
the characterictic times for mass, angular momentum and energy losses $\tau_{ml}$, $\tau_{Ml}$,$\tau_{el}$ are considerably greater then $\tau_{rel}$.  The equations describing different losses may be phenomenologically written as

\begin{equation}
\label{eq20}
\frac{dU_{tot}}{dt}=-\frac{U_{kin}}{\tau_{ml}},
\end{equation}

\begin{equation}
\label{eq21}
\frac{dm}{dt}=\frac{mU_{kin}}{U_g\tau_{ml}},\quad U_g<0,
\end{equation}

\begin{equation}
\label{eq22}
\frac{dM}{dt}=\frac{MU_{kin}}{U_g\tau_{Ml}}.
\end{equation}
Scaling all the characteristic times by the Jeans value, like in the case (\ref{eq17}) of $\tau_{rel}$, we have

\begin{equation}
\label{eq23}
\tau_{el}=\alpha_{el}\tau_J, \quad 
\quad \tau_{ml}=\alpha_{ml}\tau_J, \quad \tau_{Ml}=\alpha_{Ml}\tau_J,
\end{equation}
with constant values of $\alpha_i\,\, (i=el,\, ml,\, Ml)$.
Combining (\ref{eq21}),(\ref{eq22}), we find a relation

\begin{equation}
\label{eq24}
\frac{M}{M_{in}}=\left(\frac{m}{m_{in}}\right)^{\frac{\alpha_{ml}}
{\alpha_{Ml}}},
\end{equation}
where $m_{in}$ and $M_{in}$ are the initial values of corresponding parameters.
The function ${\cal E}$ determines the entropy of the matter in the spheroid, so that, using (\ref{eq10}), we have
 
\begin{equation}
\label{eq25}
\frac{dE_{th}}{dt}+ P\frac{dV}{dt}=\frac{1}{(abc)^{2/3}}
\frac{d{\cal E}}{dt},
\end{equation}
$$
E_{th}=\frac{\cal E}{(abc)^{2/3}}, \quad V=\frac{4\pi}{3}abc,\quad 
P=\frac{2}{3}\frac{E_{th}}{V}.
$$
Using (\ref{eq20})-(\ref{eq22}) in (\ref{eq19}) we obtain the equation for the entropy function $\cal E$ in presence of different losses 
in the form

\begin{equation}
\label{eq26}
\frac{d{\cal E}}{dt}=(abc)^{2/3}U_{kin}\biggl[
\left(\frac{2}{\tau_{rel}}-\frac{1}{\tau_{el}}-\frac{2}{\tau_{ml}}\right)-
\frac{U_{rot}}{U_g}\left(\frac{2}{\tau_{Ml}}-\frac{1}{\tau_{ml}}\right)+\frac{U_{kin}}{U_g\tau_{ml}}
\biggr].
\end{equation}

\section{Non-dimensional equations}

The equations, describing approximately the dynamics of the fomation of a stationary dark matter object include equations of motion (\ref{eq7}),(\ref{eq8}) with adding terms from (\ref{eq12}); energy equation (\ref{eq26}) with account of (\ref{eq3}),(\ref{eq10}),(\ref{eq11}); and equations (\ref{eq20})-(\ref{eq22}), describing the losses of the energy, mass and angular momentum. The mass losses are also taken into account in the equations of motion, written in the form (\ref{eq14})-(\ref{eq16}). For obtaining a numerical solution of these equations let us write them in non-dimensional variables. Introduce the following non-dimensional variables

\begin{equation}
\label{eq27}
\tilde t=\frac{t}{t_0},\,\,\tilde a=\frac{a}{a_0},\,\,\tilde c=\frac{c}{a_0},\,\,\tilde m=\frac{m}{m_0},\,\,\tilde M=\frac{M}{M_0},
\end{equation}
$$
\tilde \rho=\frac{\rho}{\rho_0},\,\,\tilde U=\frac{U}{U_0},\,\,\tilde E_{th}=\frac{E_{th}}{U_0},\,\,{\cal E}=\frac{{\cal E}}{\cal E}_0,\,\,
\tilde \tau_i=\frac{\tau_i}{t_0}.
$$
The process, with account of mass and angular momentum losses, is described by the following non-dimensional system of equations

\begin{equation}
\label{eq28}
\ddot{a}=\frac{\dot a}{m}\frac{a(2\dot a^2+\dot c^2)}{6\tau_{ml} a_k}
+\frac{25}{4}\frac{M^2}{m^2a^3}+\frac{3}{2}\frac{m}{a^2}a1_k
-\frac{\dot a}{\tau_{rel}}+\frac{10}{3am}\frac{{\cal E}}{(a^2c)^{2/3}},
\end{equation}

\begin{equation}
\label{eq29}
\ddot{c}=\frac{\dot c}{m}\frac{a(2\dot a^2+\dot c^2)}{6\tau_{ml}a_k}
-3\frac{m}{a^2}a2_k
-\frac{\dot c}{\tau_{rel}}+
\frac{10}{3cm}\frac{{\cal E}}{(a^2c)^{2/3}},
\end{equation}

\begin{equation}
\label{eq30}
\frac{d{\cal E}}{dt}=\frac{m(2{\dot a}^2+{\dot c}^2)}
{10}(a^2c)^{2/3}\biggl[
\left(\frac{2}{\tau_{rel}}-\frac{1}{\tau_{el}}-\frac{2}{\tau_{ml}}\right)
\end{equation}
$$
+\frac{25}{12}\frac{M^2}{m^3a\, a_k}
\left(\frac{2}{\tau_{Ml}}-
\frac{1}{\tau_{ml}}\right)+\frac{a(2\dot a^2+\dot c^2)}{6m\tau_{ml}a_k}
\biggr],
$$

\begin{equation}
\label{eq31}
\frac{dm}{dt}=-\frac{a(2{\dot a}^2+{\dot c}^2)}{6\tau_{ml}a_k}
,\quad
\frac{M}{M_{in}}=
\left(\frac{m}{m_{in}}\right)^{\frac{\alpha_{ml}}{\alpha_{Ml}}}, \quad
\Omega=\frac{5}{2}\frac{M}{ma^2},
\end{equation}
\begin{equation}
\label{eq31f}
\frac{df}{dt}=\Omega.
\end{equation}
Here 
\begin{equation}
\label{eq31o}
a_k=\frac{\arccos{k}}{\sqrt{1-k^2}}, \quad a1_k=\frac{k-a_k}{1-k^2},\quad
a2_k=\frac{1-k\,a_k}{1-k^2}
\end{equation}
for the oblate spheriod $a=b>c$, and 
\begin{equation}
\label{eq31p}
a_k=\frac{{\rm Arch}{k}}{\sqrt{k^2-1}}, \quad a1_k=-\frac{k-a_k}{k^2-1},\quad
a2_k=-\frac{1-k\,a_k}{k^2-1} 
\end{equation}
for the prolate spheriod with $a=b<c$. 
In both cases $a_k+a1_k=k\, a2_k$. The rotation phase $f$ is calculated in 
(\ref{eq31f}) for comparison of the rotation and  oscillation processes.
When the spheroid is close to the sphere
$|k-1| \ll 1$ we have the following expansions 
\begin{equation}
\label{eq31op}
a_k\approx \frac{4}{3k}-\frac{1}{3k^3},\,\,
a1_k\approx-\frac{1}{k}+\frac{1}{3k^3},\,\, 
a2_k\approx\frac{1}{3k^2}\quad {\rm at}\,\,\, a=b>c,
\end{equation}
$$
a_k\approx \frac{7}{6}-\frac{k^2}{6},\,\,
a1_k\approx-\frac{1}{k+1}-\frac{1}{6},\,\, 
a2_k\approx\frac{1}{k+1}-\frac{k}{6}\quad {\rm at}\,\,\, a=b<c.
$$
In the case of a very compressed spheroid (pancake) we have the following expansion

\begin{equation}
\label{eq31pan}
a_k=\frac{\pi}{2}-k+\frac{\pi}{4}k^2-\frac{2}{3}k^3,\,\,\,
a1_k=-\frac{\pi}{2}+2k-\frac{3\pi}{4}k^2,\,\,\,
a2_k=1-\frac{\pi}{2}k+2k^2.
\end{equation}
In equations (\ref{eq28})-(\ref{eq31f}) only non-dimentional variables are used, and "tilde" sign is omitted for simplicity. The non-dimensional relaxation times $\tau_i$ in these equations are written with account of (\ref{eq18}),(\ref{eq23}) as

\begin{equation}
\label{eq32}
\tau_i=2 \pi \alpha_i\sqrt{\frac{a^2 c}{3m}}.
\end{equation}
The scaling parameters 

$$ {t_0},\,\,{a_0},\,\,{m_0},\,\,{M_0},\,\,{\rho_0},\,\,{U_0},\,\,\Omega_0,\,\,
{\cal E}_0
$$
are connected by the following relations

\begin{equation}
\label{eq33}
t_0^2=\frac{a_0^3}{Gm_0},\,\, M_0^2=Ga_0m_0^3,\,\, U_0=\frac{Gm_0^2}{a_0},
\,\, \rho_0=\frac{m_0}{a_0^3},\,\, \Omega_0=\frac{M_0}{m_0a_0^2},\,\,{\cal E}_0=U_0a_0^2,
\end{equation} 
so that in non-dimensional variables $m=4\pi a^2 c\rho/3$, $M=2ma^2\Omega/5$.
The total nondimensional energy $\tilde U_{tot}=U_{tot}/U_0$ is written as ("tilde" is omitted)

\begin{equation}
\label{eq33u}
U_{tot}=\frac{m(2{\dot a}^2+{\dot c}^2)}{10}-\frac{3m^2}{5a} a_k+\frac{5}{4}\frac{M^2}{ma^2}+\frac{\cal E}{(a^2c)^{2/3}}
\end{equation}

\section{Equilibrium configuration and linear oscillations}

In equilibrium there is no dissipation and the following relations follow from 
(\ref{eq28}),(\ref{eq29}), connecting the equilibrium parameters

\begin{equation}
\label{eq28eq}
\frac{25}{4}\frac{M^2}{m^2a^3}+\frac{3}{2}\frac{m}{a^2}\,a1_k\,
+\frac{10}{3am}\frac{{\cal E}}{(a^2c)^{2/3}}=0,
\end{equation}

\begin{equation}
\label{eq29eq}
-3\frac{m}{a^2}\,a2_k\,+
\frac{10}{3cm}\frac{{\cal E}}{(a^2c)^{2/3}}=0.
\end{equation}
At given $m,M$ and ${\cal E}$ these equations determine the con\-fi\-gu\-ra\-ti\-on of the sphe\-ro\-id. Finding $a$ from (\ref{eq29eq})

\begin{equation}
\label{eq30eq}
a=\frac{10}{9m^2}\frac{{\cal E}}{k^{5/3}\,a2_k},
\end{equation}
we obtain after its substitution into (\ref{eq28eq}) the relation for determining $k$ in the form

\begin{equation}
\label{eq31eq}
a1_k\,+2k\, a2_k\,+\frac{15}{4}\frac{M^2}{m}\frac{k^{5/3}\,a2_k\,}{\cal E}=0.
\end{equation}
Taking into account (\ref{eq31o}), we come to the relation  

\begin{equation}
\label{eq32eq}
\frac{k(1+2k^2)}{(1-k^2)^{3/2}}\arccos{k}-\frac{3k^2}{1-k^2}=
\frac{75}{4}\frac{M^2}{m{\cal E}}k^{5/3}\,a2_k
=\frac{25}{6}\frac{M^2}{m^3}\frac{k}{a}. 
\end{equation}
The last relation in (\ref{eq32eq}) follows from the connection (\ref{eq30eq}) between $a$ and ${\cal E}$, and in this form (\ref{eq32eq}) coinsids exactly with the equation (\ref{eq6}), determining the form of the Maclaurin spheroid with

$$a=\left(\frac{3m}{4\pi k \rho}\right)^{1/3},$$
of a given mass, density and angular momentum. In our case the density is determined by the entropy ${\cal E}$. For the spheroid with unit non-dimensional mass $m$ and semiaxis $a$ we obtain in equilibrium relations determining $M$ and ${\cal E}$ as functions of $k$

\begin{equation}
\label{eq33eq}
{\cal E}=\frac{9}{10}k^{5/3}\,a2_k, \quad
M^2= -\frac{6}{25}(a1_k\,+2k\, a2_k).
\end{equation}
The spheriod, described by ordinary differential equations for semiaxies $a$ and $c$ has two levels of freedom in oscilations corresponding to the first $p$- mode, and the first fundamental $f$-mode (see Cox, 1980). Consider (without loosing a generality) oscillations of a spheroid with $m=a=1$. The ${\cal E},U_{tot}, m$ and $M$ losses are quadratic to perturbations, so these values are conserved during linear oscillations. For the perturbations of $\delta a$ and $\delta c$ we have the following equations from (\ref{eq28}),(\ref{eq29}) 

\begin{equation}
\label{eq34o}
\ddot{\delta a}=-\frac{75}{4}M^2 \delta a+
\frac{3}{2}\delta(a1_k)-3 \,a1_k\,\delta a -10\frac{{\cal E}}{k^{2/3}} \delta a -\frac{20}{9}\frac{{\cal E}}{k^{5/3}} \delta k-\frac{{\dot\delta a}}{\tau_{rel}},
\end{equation}

\begin{equation}
\label{eq35o}
\ddot{\delta c}=-3\delta(a2_k)+6\, a2_k\,\delta a -10\frac{{\cal E}}{k^{5/3}} \delta a -\frac{50}{9}\frac{{\cal E}}{k^{8/3}} \delta k -
\frac{\dot{\delta c}}{\tau_{rel}}.
\end{equation}
The perturbations are expressed via perturbations $ \delta a,\,\,\delta c$ as

\begin{equation}
\label{eq36o}
\delta (a_k)=-a2_k\, \delta k,\,\,\delta (a1_k)=\frac{1+\,a2_k\,+2k\,a1_k\,}{1-k^2}\delta k, \quad 
\end{equation}
$$\delta (a2_k)=\frac{3k\, a2_k\,-a_k\,}{1-k^2}\delta k,\quad
\delta k=\delta c -\delta a;\,\,\,\,
\tau_{rel}=2\pi \alpha_{rel}\sqrt{\frac{k}{3}}.$$
The equations (\ref{eq34o}),(\ref{eq35o}) with account of (\ref{eq36o}) may be solved analitically at weak damping
for any $k$. Consider here for simplicity two limiting cases
$k=1$ (sphere), and $k\ll 1$ (pancake). The sphere is related to the nonrotating body with $M=0$. In this case we have also

\begin{equation}
\label{eq37o}
 a_k=1,\,\,\, a1_k=-\frac{2}{3},\,\,\,
a2_k=\frac{1}{3},\,\,\,{\cal E}=\frac{3}{10},
\end{equation}
$$\delta(a_k)=-\frac{\delta k}{3},\,\,\,\delta(a1_k)=\frac{2\delta k}{5},\,\,\,
\delta(a2_k)=-\frac{4\delta k}{15}.$$
With account of (\ref{eq37o}) we have from (\ref{eq34o}),(\ref{eq35o}) the following equations

\begin{equation}
\label{eq38o}
\ddot{\delta a}=-\frac{1}{15}\delta k - \delta a
-\frac{\dot{\delta a}}{\tau_{rel}},\,\,\,
\ddot{\delta c}=-\frac{13}{15}\delta k - \delta a
-\frac{\dot{\delta c}}{\tau_{rel}}.
\end{equation}
These equations are completed by relation $\delta k=\delta c-\delta a$. 
Taking $\delta a,\,\,\delta c$ $\sim \, \exp(-i\omega t)$
we come to the characteristic equation

\begin{equation}
\label{eq39o}
\omega^4-\frac{9}{5}\omega^2+\frac{4}{5}+\frac{i\omega}{\tau_{rel}}
\left(2\omega^2-\frac{27}{15}\right)-\frac{\omega^2}{\tau_{rel}^2}=0
\end{equation}
The solution of a characteristic equation may be obtained analytically for a weak damping at $\omega \gg 1/\tau_{rel}$. The 4 roots of this equation are

\begin{equation}
\label{eq40o}
\omega_{1,2}= \pm 1-\frac{i}{2\tau_{rel}}, \quad
\omega_{3,4}= \pm \sqrt{\frac{4}{5}}-\frac{i}{2\tau_{rel}} 
\end{equation}
The first root corresponds to a pure radial oscillations, and the second one describes the fundamental mode which does not change the volume and which is preserved in the non-compressible fluid. In fact, for a pure radial oscillations with $\delta k=0$ we have $\delta c=\delta a$, and the equations (\ref{eq38o}) reduce to $\ddot{\delta a}=
-\delta a-\frac{\dot{\delta a}}{\tau_{rel}}$ with a characteristic equation
$\omega^2+\frac{i\omega}{\tau_{rel}}-1=0$. The solution of this equation is
$\omega_{1,2}=-\frac{i}{2\tau_{rel}} \pm \sqrt{1-\frac{1}{4\tau_{rel}^2}}$,
which for a small damping obviously correspond to the first two roots in (\ref{eq40o}). The damping increments are equal for these two modes, what may be seen also from the numerical solution of (\ref{eq28})-(\ref{eq31f})
for a general nonstationary problem at large $t$ (see next section).
Note, that in the radial oscillation mode $\delta a =\delta c$, and in the fundamental mode $\delta a \approx -0.5\delta c$. 

For the case of a pancake $k\ll 1$ we have from  (\ref{eq31pan}),(\ref{eq33eq}),
(\ref{eq36o})

\begin{equation}
\label{eq41o}
 a_k=\frac{\pi}{2},\,\,\, a1_k=-\frac{\pi}{2},\,\,\,
a2_k=1,\,\,\,{\cal E}=\frac{9}{10}k^{5/3},\,\,\, M^2=\frac{3\pi}{25},
\end{equation}
$$\delta(a_k)=-\delta k,\,\,\,\delta(a1_k)=2\delta k,\,\,\,
\delta(a2_k)=-\frac{\pi \delta k}{2}.$$
With account of (\ref{eq41o}) we have from (\ref{eq34o}),(\ref{eq35o}) the following equations

\begin{equation}
\label{eq42o}
\ddot{\delta a}=-\frac{3\pi}{4}\delta a - 3\delta k
-\frac{\dot{\delta a}}{\tau_{rel}},\,\,\,
\ddot{\delta c}=-\frac{5}{k}\delta k - 3\delta a
-\frac{\dot{\delta c}}{\tau_{rel}},
\end{equation}
with $\delta k=\delta c-k\delta a$.
Taking $\delta a,\,\,\delta c$ $\sim \, \exp(-i\omega t)$
we come to the characteristic equation

\begin{equation}
\label{eq43o}
\omega^4-\left(\frac{5}{k}+\frac{3\pi}{4}\right) \omega^2+\frac{15\pi}{4}+6+\frac{i\omega}{\tau_{rel}}
\left(2\omega^2-\frac{5}{k}-\frac{3\pi}{4}\right)-
\frac{\omega^2}{\tau_{rel}^2}=0
\end{equation}
The solution of a characteristic equation may be obtained analytically for a weak damping at $\omega \gg 1/\tau_{rel}$. The 4 roots of this equation are

\begin{equation}
\label{eq44o}
\omega_{1,2}= \pm \sqrt{\frac{5}{k}}-\frac{i}{2\tau_{rel}}, \quad
\omega_{3,4}= \pm \sqrt{\frac{3\pi}{4}}-\frac{i}{2\tau_{rel}} 
\end{equation}
In the pancake the oscillation modes (1,2) and (3,4) are almost independent on each other. In the first two modes $\delta a\sim k\delta c \ll \delta c$, what corresponds to rapid oscillations of the thickness of the pancake at almost fixed radius. Two other modes with a low frequency are characterized by relation $\delta c\sim k\delta a \ll \delta a$, and correspond to the oscillation of the equatorial radius of the pancake at approximately the same thickness. Here both modes have the same damping increments because we have chosen the same relaxation time for the damping of two degrees of freedom, which may differ in reality for the flat pancake. 

The model, developed here may be easily generalized onto the 3 dimensional ellipsoids, where instead of analytical formulae for the gravitational potential and forces we'll have elliptical integrals. This model will have three degrees of freedom, with the additional one, corresponding to non-axisymmetric perturbations. Contrary to the axisymmetric modes, which are always stable, the  non-axisymmetric mode of the spheroid will become unstable at sufficiently large $M$, first due to a secular instability appearing only at presence of damping, and at larger $M$ the spheriod will become dynamically unstable independently on the presence of the damping. In that case we may expect that in presence of damping the dynamical instability will appear at larger $M$, than in the pure conservative case (see Chandrasekhar, 1969). We suppose to investigate the three-dimensional model and its stability in the near future.

\section{Numerical results}

The development of gravitational instability starts from almost Hubble expansion velocity which is transformed into contraction as a result of a developement of gravitational instability. To simplify the problem we start from a nonlinear phase of the instability with an expansion velocity smaller than the Hubble value. We start the simulation from the spherical body of the unit mass, small or zero entropy ${\cal E}_{in}\ll 1$, and negative total energy
$U_{tot}<0$. We also specify the parameters, 
  characterizing different dissipations
$\alpha_{rel},\,\,\,\alpha_{el},\,\,\, \alpha_{ml},\,\,\, \alpha_{Ml}$. The following parameters have been used in all variants of calculations
\begin{equation}
\label{eq45r}
a_{in}=1,\,\,\,c_{in}=1,\,\,\,\,\,\, m_{in}=1,\,\,\,\alpha_{rel}=3,\,\,\,\alpha_{el}= \alpha_{ml}=\alpha_{Ml}=15. 
\end{equation}
Other four initial parameters $\dot a_{in},\,\, \dot c_{in},\,\, M$ and 
${\cal E}$
used in calculations are listed in the Table 1, where also the initial $U_{tot}$ is represented, which is calculated from (\ref{eq19}).

\begin{table}[h]
\vspace*{0.5cm}
\begin{center}
\begin{tabular}{|c|c|c|c|c|c|}
\hline
  N   &$\dot a_{in}$ &$\dot c_{in}$ &   $ M $      & ${\cal E}$ &$U_{tot}$ \\
\hline
  1   &    0.5       &   0.5        &    0.3       &    0.01    & -0.4025 \\
  2   &    0.5       &   0.5        &    0.5       &    0.01    & -0.2025 \\
  3   &    0.2       &   0.2        &    0.5       &    0.01    & -0.2655 \\
  4   &    0.5       &   0.5        &    0.1       &    0       & -0.5125 \\
  5   &    0.9       &   0.9        &    0.1       &    0       & -0.3445 \\
  6   &    0.9       &   0.7        &    0.1       &    0       & -0.3765 \\
  7   &    0.99      &   0.94       &    0.01      &    0       & -0.3155 \\
  8   &    1.3       &   1.1        &    0.01      &    0       & -0.1409 \\
 \hline
\end{tabular}
\caption{Initial conditions and parameters taken in numerical solution of equations (\ref{eq28})-(\ref{eq31f}), in addition to the fixed values from (\ref{eq45r}); and initial value of the total energy, according to (\ref{eq19}).}
\label{initial} 
\end{center}
\end{table}

We have used in the calculations the Runge-Kutta code from (Press et al., 1992).
The results of the calculations are very transparent, and are represented in figures 1-16. The odd number figures represent time dependences of $a$ and $c$ axies on time, together with the $phase=\sin(ph)=\sin(\int_0^t{\Omega dt})$, permitting the comparison between the oscillational and rotational motion. In the even number figures the entropy ${\cal E}$, current mass $m$ and current total energy $U_{tot}$ are given as functions of the time for the choosen set of the relaxation parameters  (\ref{eq45r}). Here the entropy behaviour is the most interesting, because it is based on the realistic estimation of $\tau_{rel}$. The characteristic times of other losses are more difficult to estimate, so they had been choosen equal to $5\tau_{rel}$, making these losses practically unimportant.
  
 After the initial expansion the collapse happens, during which the linear size decreases 10-40 times, and in the process of the subsequent relaxation the object approaches the equilibrium, described by relations (\ref{eq33eq}). The relaxation and entropy production, mass and energy losses happens in the most rapid stages at maximal compression, what is reflected in the steps of the functions ${\cal E}(t),\, m(t),\, U_{tot}(t)\,$, well visible in figures 8,10,12,14,16. The following remarkable properties should be noted.

{\bf 1.} For a low initial angular momentum, which is probably characteristic for the collapse of the primodial dark matter clouds, the collapse leads initially to the "pancake", in accordance of Zeldovich (1970) theory. On the subsequent stage of the expansion the spheroid becomes prolate, and during subsequent phases the spheroid is oscillating on the fundamental mode, changing the form between the prolate and oblate ones. It may be seen especially well in figures 11 and 15, where temporally the prolate figure is formed with $c/a \sim 2-5$. We may expect the formation of such transient prolate figures in the central parts of the collapsing dark matter cloud, where the density may be taken as almost uniform, and where the characteristic period of oscillations is much less than the time of the whole body collapse.

{\bf 2.} The phenomenological relaxation time accepted in the calculations seems to be rather realistic (Lynden-Bell, 1967), so the calculated process of the damping should be close to a reality. We have obtained, that oscillations preserve during more than 10 initial characteristic (Jeans) times $t_J$. That means, that in the most extended dark matter objects with the largest $t_J$ the damping of oscillations is still in progress, and the dark matter objects may be presently in a nonstationary state. 
The collapsing dark matter ofjects and their oscillations 
 could influence the visible picture of the fluctuations of the cosmic background microwave radiation, which are now under intensive investigation (see Naselsky, Novikov, Novikov, 2002).

{\bf 3.} The description of the dynamics of the spheroid by the system of equations with two degrees of freedom give a possibility of the exitation of only two oscillation modes. In the case of a low rotation when two degrees of freedom are strongly coupled the two modes reduce to the first radial $p$-mode, like the main mode of the radial oscillations of the sphere, and the fundamental $f$-mode, which almost does not change the total volume, and which survives in the non-compressible sphere (Cox, 1980). The lengths of two axies oscillate with the same period. Excitation of both modes may be distinctly seen in calculations. For some initial conditions the main remaining mode is the $p$-mode (figures 7,15); for others  (figures 9,11,13) the fundamental $f$-mode was mainly excited.

For rapidly rotating bodies, with a large equilibrium ratio of $a/c$, the oscillations in two axies are almost independent, and their lengths are oscillating with substantially different periods (figures 1,3,5).        

\section{Emission of very long graviattional waves}

Gravitational radiation is produced during the collapse of the nonspherical body. Gravitational radiation during a formation of a pancake was estimated by Thuan and Ostriker (1974), and was improved by Novikov (1975), who took into account the most important stage of the radiation during a bounce. The formula for the estimation of the total energy emitted during the collapse and bounce is

\begin{equation}
\label{eq46w} 
  U_{GW}\approx 0.01 \left(\frac{r_g}{a_{eq}}\right)^{7/2}
 \left(\frac{a_{eq}}{c_{min}}\right)\,Mc^2.
\end{equation}
Here $a_{eq}$ is the equilibrium equatorial radius of the pancake, and $c_{min}$ is its minimal half-thickness during the bounce, $r_g=2GM/c^2$ is the Schwarzschild gravitational radius of the body. 
 From our calculations we have $a_{eq}/c_{min} \le 100$. The value of   ${r_g}/{a_{eq}}$ we estimate using the observed average velocity of a galaxy in  the rich cluster $v_g \sim 3000$ km/s, and taking

\begin{equation}
\label{eq47w}
\frac{r_g}{a_{eq}} \sim \left(\frac{v_g}{c}\right)^2 \approx 10^{-4}.
\end{equation}
Than the energy carried away by the gravitational wave (GW) may be estimated as
$U_{GW} \approx 10^{-14} M\,c^2$. If all dark matter had passed through the stage of a pancake formation, than very long GW with a wavelength of the order of the size of the galactic cluster have an average energy density in the universe 

\begin{equation}
\label{eq48w}
\varepsilon_{GW} \approx 10^{-14} \rho_{dm} c^2\approx 3\cdot 10^{-23} {\rm erg/cm}^3.
\end{equation}  
Here we have used for estimation the average dark matter density $\rho_{dm}=3\cdot 10^{-30}$ g/cm$^3$. 
The average strength $E_{GW}$ of the very long GW may be estimated, taking
the relation (Landau and Lifshits, 1993)

\begin{equation}
\label{eq49w}
\varepsilon_{GW}=\frac{c^2}{16\pi G}{\dot h}^2,
\end{equation}
where $ h $ is metric tensor perturbation (nondimensional), connected with GW, we consider only the scalar, having in mind the averaged value of this perturbation. Taking into account ${\dot h}=\omega h=2\pi c h/\lambda$,\,  $\lambda\sim 10$ Mpc is the wavelength of GW of the order of the size of the cluster. From the comparison of  (\ref{eq48w}),(\ref{eq49w}) we obtain the averaged metric perturbation due to very long GW in the form

\begin{equation}
\label{eq50w}
\bar h=\frac{2\lambda}{c^2}\left(\frac{G\varepsilon_{GW}}{\pi}\right)^{1/2} \approx 6\cdot 10^{-11}
\end{equation}
for the values of $\lambda$ and $\varepsilon_{GW}$, mentioned above. Insidently we may expect 10 times larger amplitude of the GW, than the averaged over the volume value. We may estimate the angle of the lensing using its linear dependence on the potential difference, and the value of deviation $1.75$ arc sec. for the Sun (Zeldovich and Novikov, 1971) which gravitational potential is equal to $h_{sun}=GM/Rc^2 \approx 2 \cdot 10^{-6}$. Finally we get for the expected lensing influence of the very long GW

\begin{equation}
\label{eq51w}
\theta_{GW}=\theta_{sun}\frac{10\bar h}{h_{sun}}= 3\cdot 10^{-4}\theta_{sun}\approx 5\cdot 10^{-4} {\rm arc sec.}
\end{equation}
 Such angular resolution is available by the very long base interferometry
and the interferometric optical telescopes under construction could give much 
better angular resolution.

\section{Conclusions}

The simplified model of the collapse and subsequent relaxation of the rotating compressed shperoid may be related to the processes in the central parts ob the dark matter objects. The formation of the transient prolate dark matter spheroid follows from the calculations for slowly rotating (or non-rotating) objects. 

For realistic values of the relaxation time we have obtained rather slow damping of the oscillations, indicating that after $\sim 10$ oscillations the amplitude may remain on the level of 1/10 of the Hubble value of the velosity for a corresponding mass scale. Such oscillations in the dark matter objects of the largest scale may be preserved until the present time.

Interaction of the cosmic microvawe background radiation (CMB) with the dark matter objects on the stage of their collapse and subsequent oscillation phases 
may influence the visible characteristics of the observed picture of the fluctuations of CMB.   

The weak but very long gravitationl waves (GW) emitted mainly on the stages of the collapse and pancake formation form a long wave GW background, which also influence CMB fluctuations (Doroshkevich et al, 1977).

 Besides, the existence of such a long GW in the space between the source and the observer may be registered as an action of the gravitational lense. Absence of the appropriate lensing objects in the direction of any observed gravitational lense object may indicate to the existence of dark matter objects without the barion matter presence, or may be the indication to the presence of the very long GW which is moving in the space between us and the lensed object The expected angles of such lensing ($\sim 10^{-3}$) arc sec. may be available in the near future. 

The present approach describing the compressed spheroid by the two degrees of freedom may be evidently generalized for a description of the 3-axis spheroid by the ordinary equations with 3 degrees of freedom. In particular, it would be possible to investigate the influence of the damping processes on the boundaries of the stability of the Maclaurin spheroid for the transformation into the Jacobi 3-axis ellipsoid (secular and dynamical) at large angular momentums. 
The model could be generalized also for the case of anizotropic relaxation and creation of anizotropic pressure.

\begin{figure}
\centerline
{
\psfig{figure=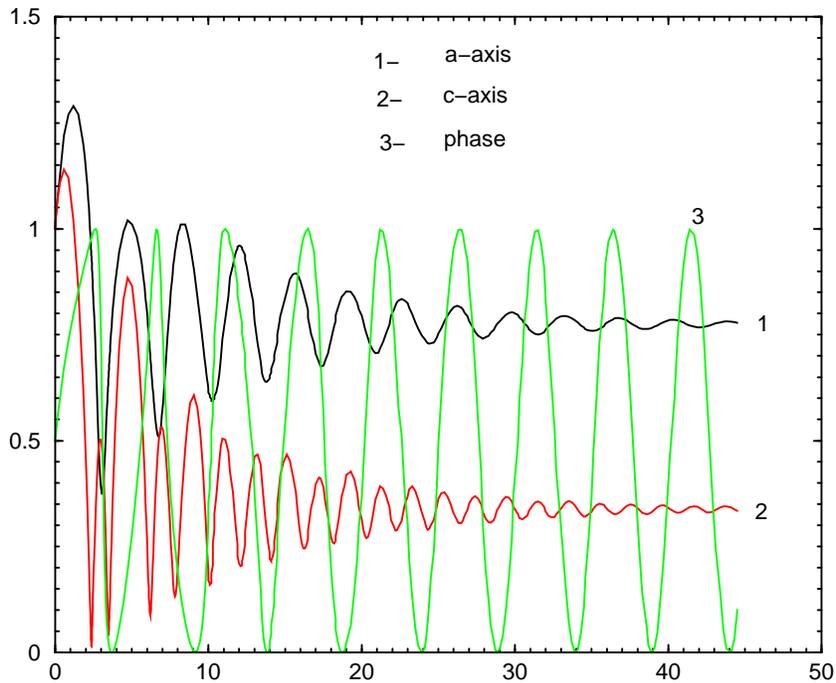,width=11cm,angle=-90}
}
\caption{Time dependence of semiaxies $a(t)$ and $c(t)$, and sinus of the rotational phase: $phase(t)$ for initial values of the variant 1, Table 1. }
\label{fig1}
\end{figure}

\begin{figure}
\centerline{
\psfig{file=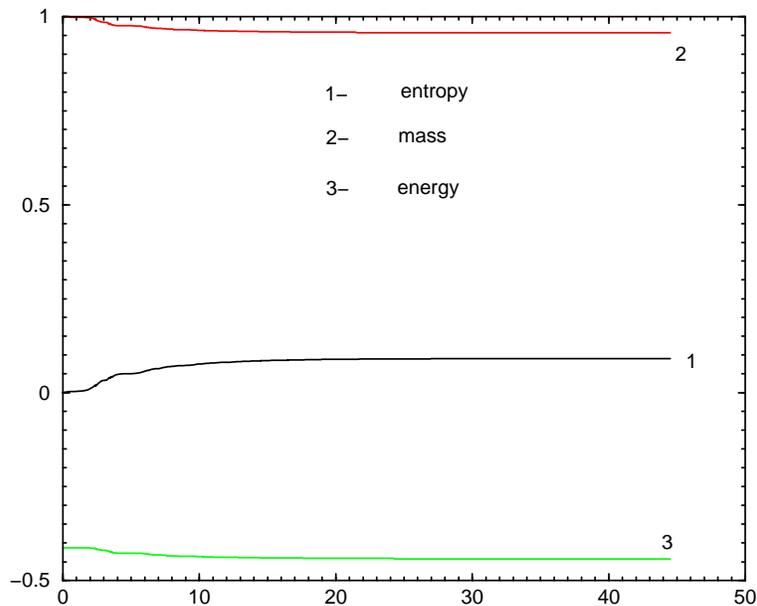,width=10cm,angle=-90}
}
\caption{Time dependence of the non-dimensional entropy ${\cal E}(t)$, mass $m(t)$, and total energy $U_{tot}(t)$ for initial values of the variant 1, Table 1.} 
\label{fig1a}
\end{figure}

\begin{figure}
\centerline
{
\psfig{figure=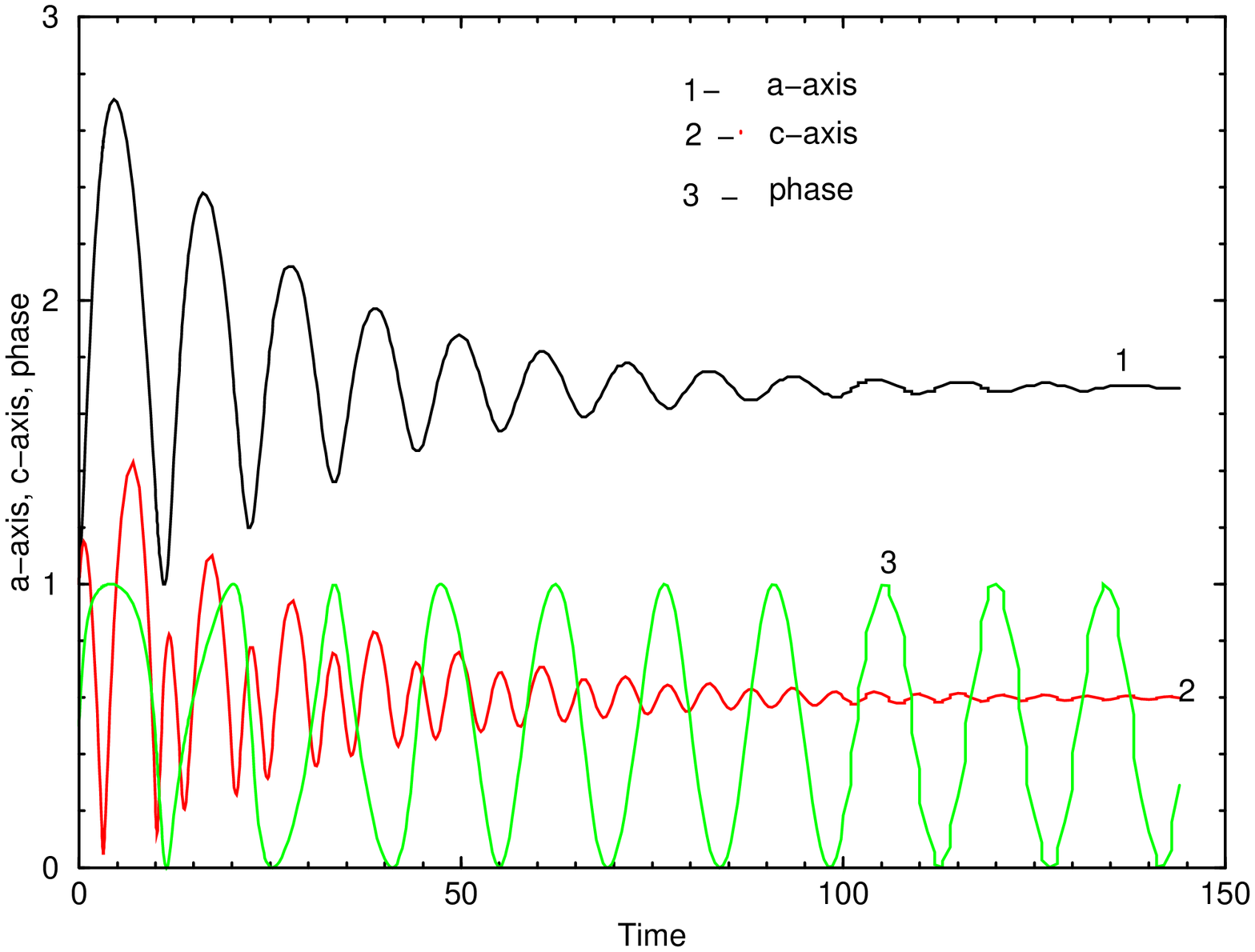,width=11cm,angle=-90}
}
\caption{Same as in Fig.1, for initial values of the variant 2, Table 1. }
\label{fig2}
\end{figure}

\begin{figure}
\centerline{
\psfig{file=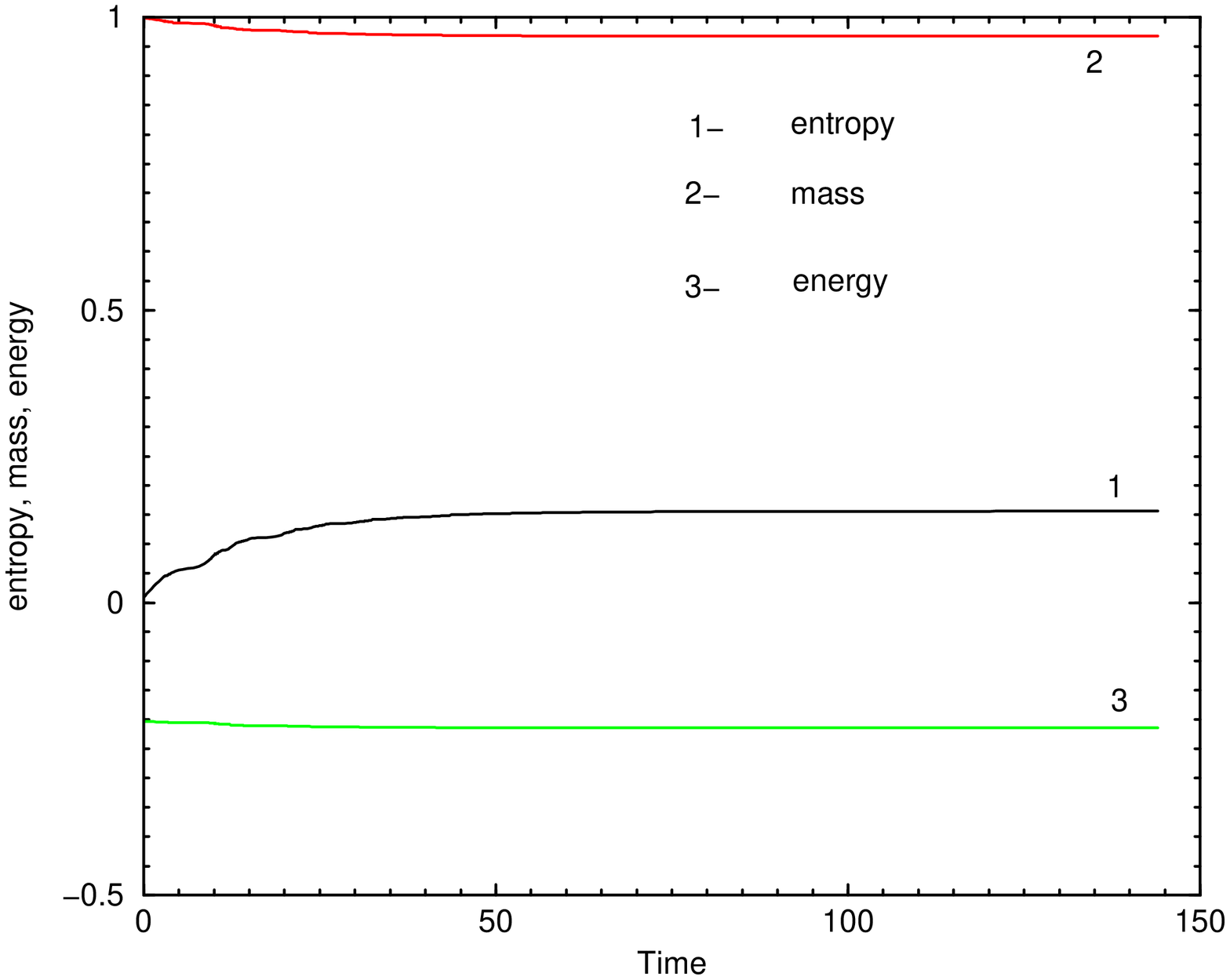,width=10cm,angle=-90}
}
\caption{Same as in Fig.2, for initial values of the variant 2, Table 1.} 
\label{fig2a}
\end{figure}

\begin{figure}
\centerline
{
\psfig{figure=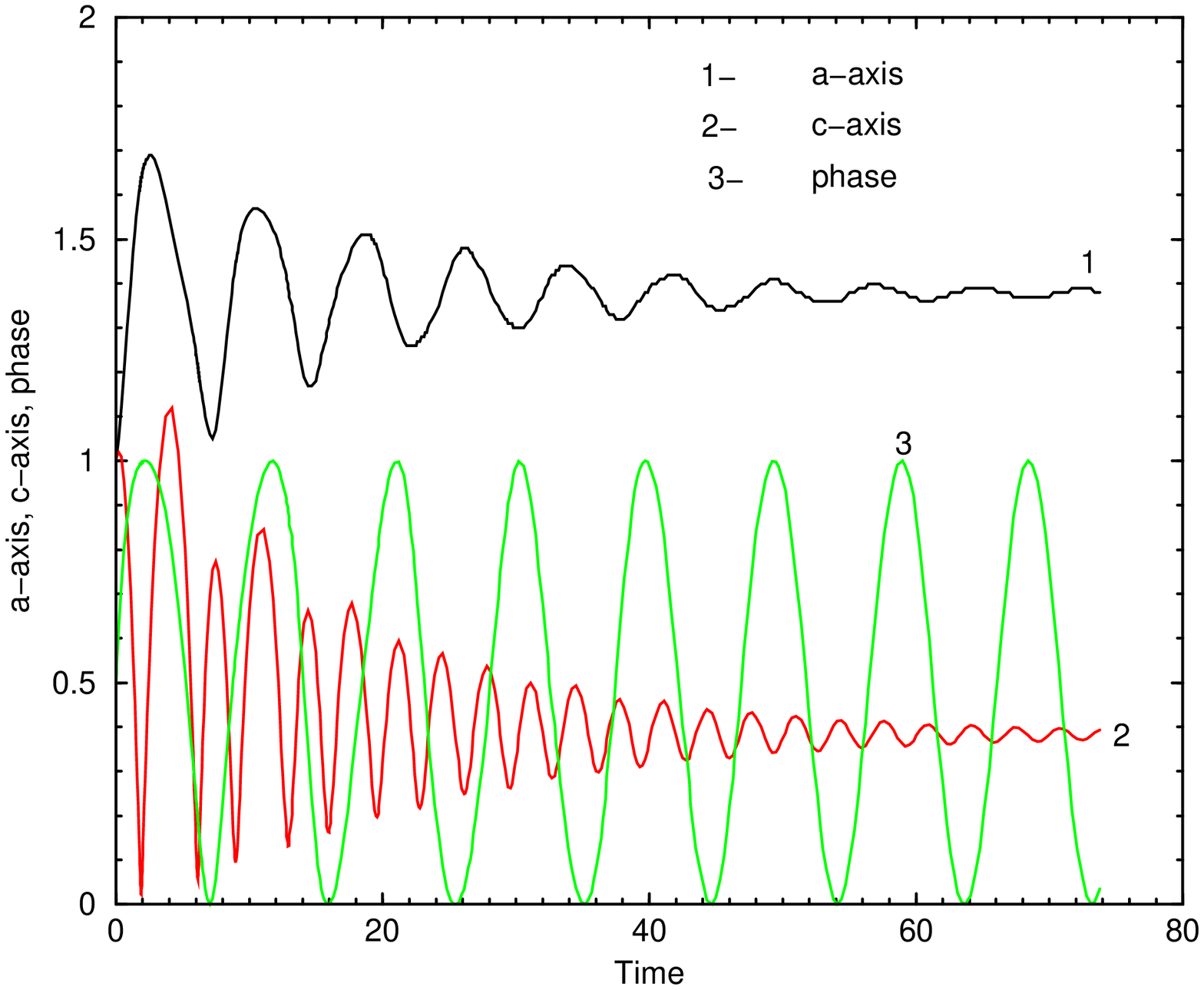,width=11cm,angle=-90}
}
\caption{Same as in Fig.1, for initial values of the variant 3, Table 1. }
\label{fig3}
\end{figure}

\begin{figure}
\centerline{
\psfig{file=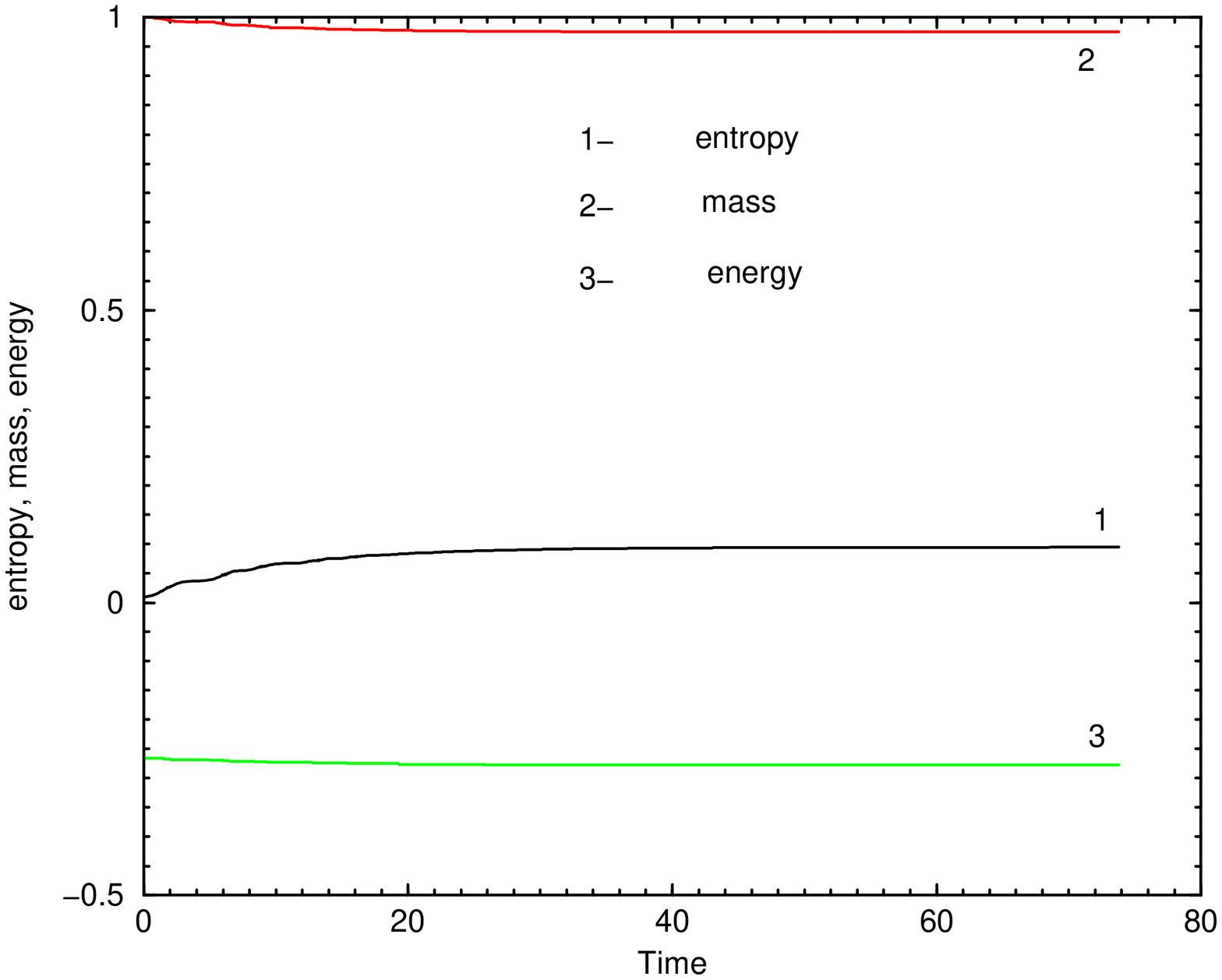,width=10cm,angle=-90}
}
\caption{Same as in Fig.2, for initial values of the variant 3, Table 1.} 
\label{fig3a}
\end{figure}

\begin{figure}
\centerline
{
\psfig{figure=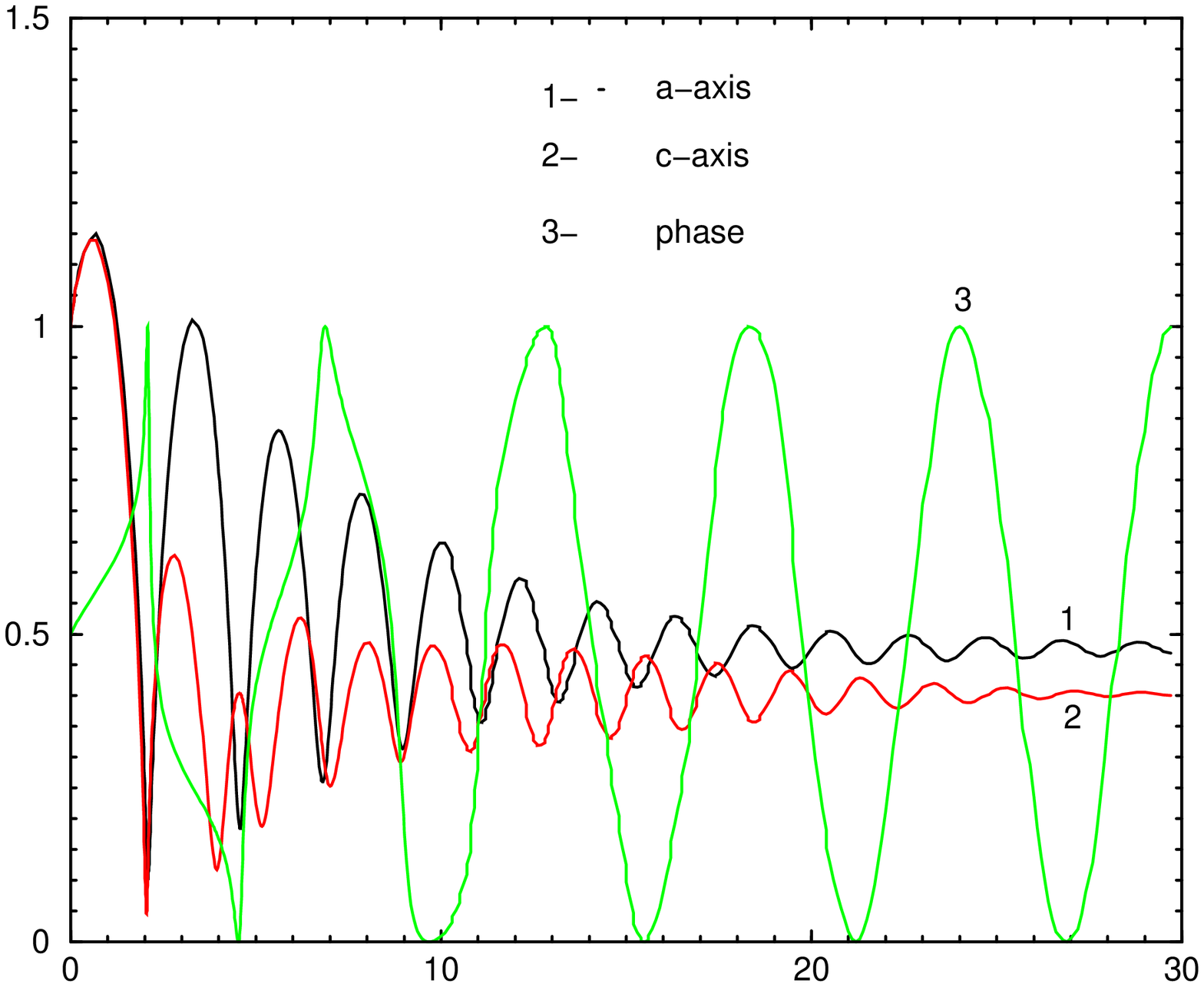,width=11cm,angle=-90}
}
\caption{Same as in Fig.1, for initial values of the variant 4, Table 1. }
\label{fig4}
\end{figure}

\begin{figure}
\centerline{
\psfig{file=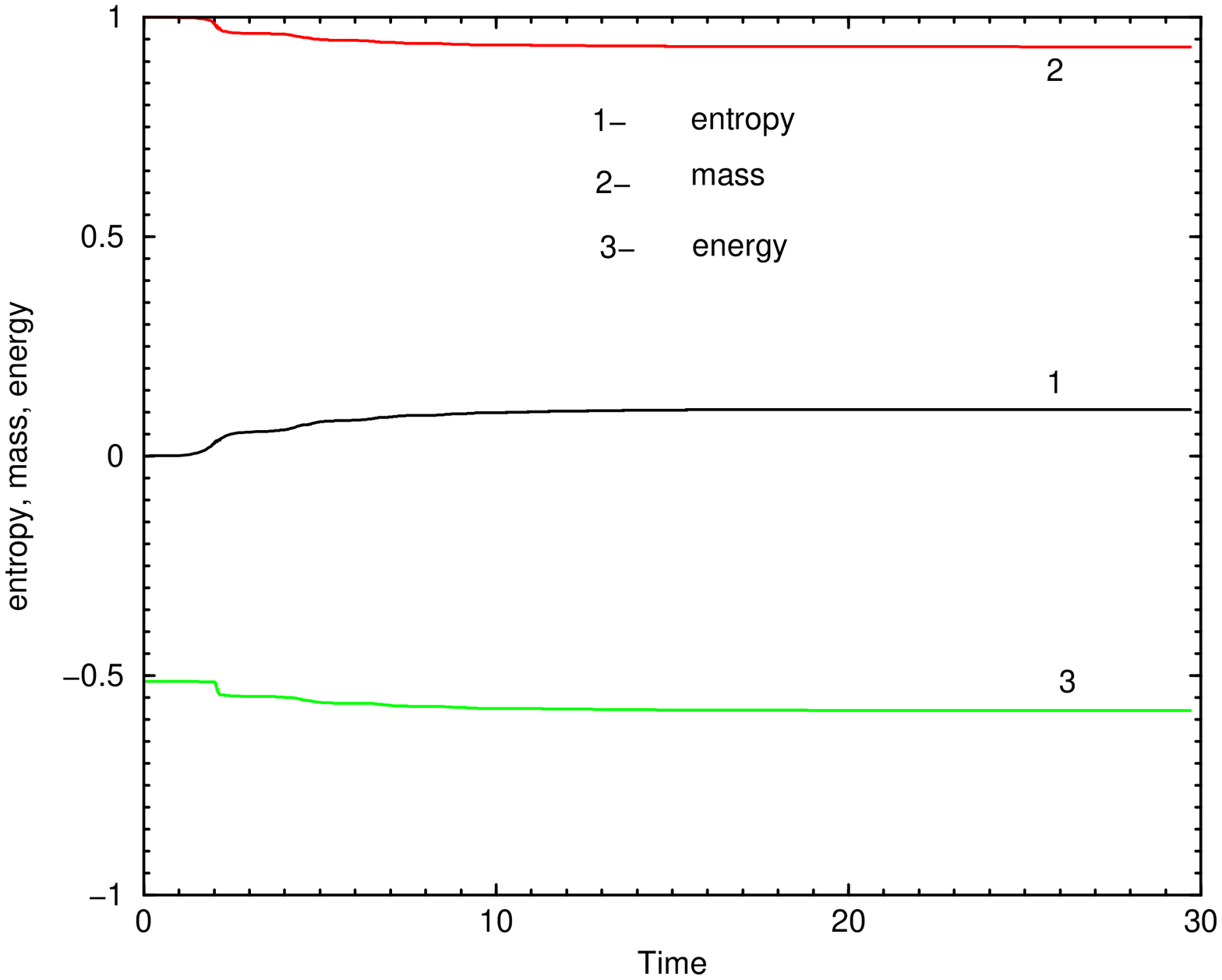,width=10cm,angle=-90}
}
\caption{Same as in Fig.2, for initial values of the variant 4, Table 1.} 
\label{fig4a}
\end{figure}

\begin{figure}
\centerline
{
\psfig{figure=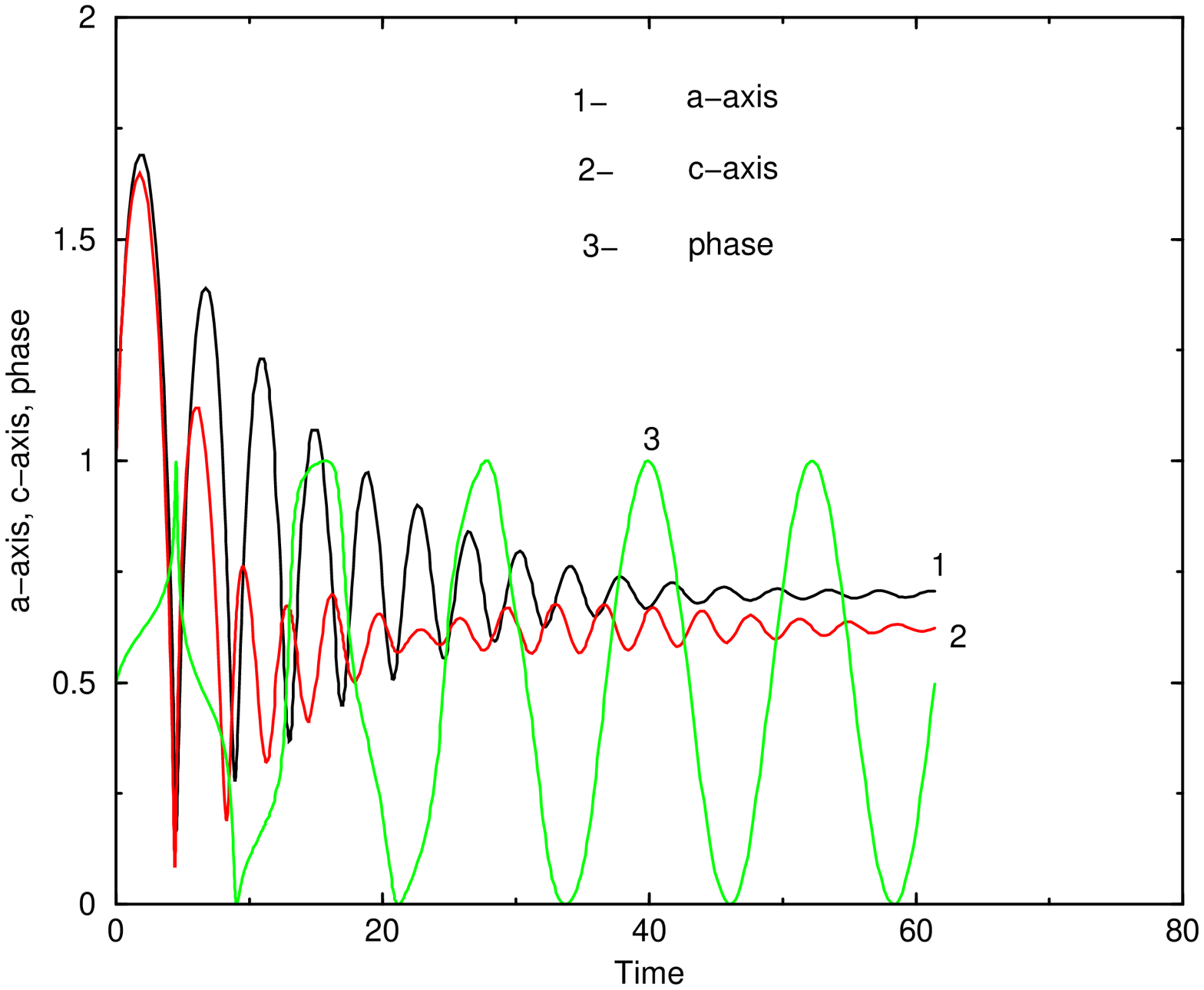,width=11cm,angle=-90}
}
\caption{Same as in Fig.1, for initial values of the variant 5, Table 1. }
\label{fig5}
\end{figure}

\begin{figure}
\centerline{
\psfig{file=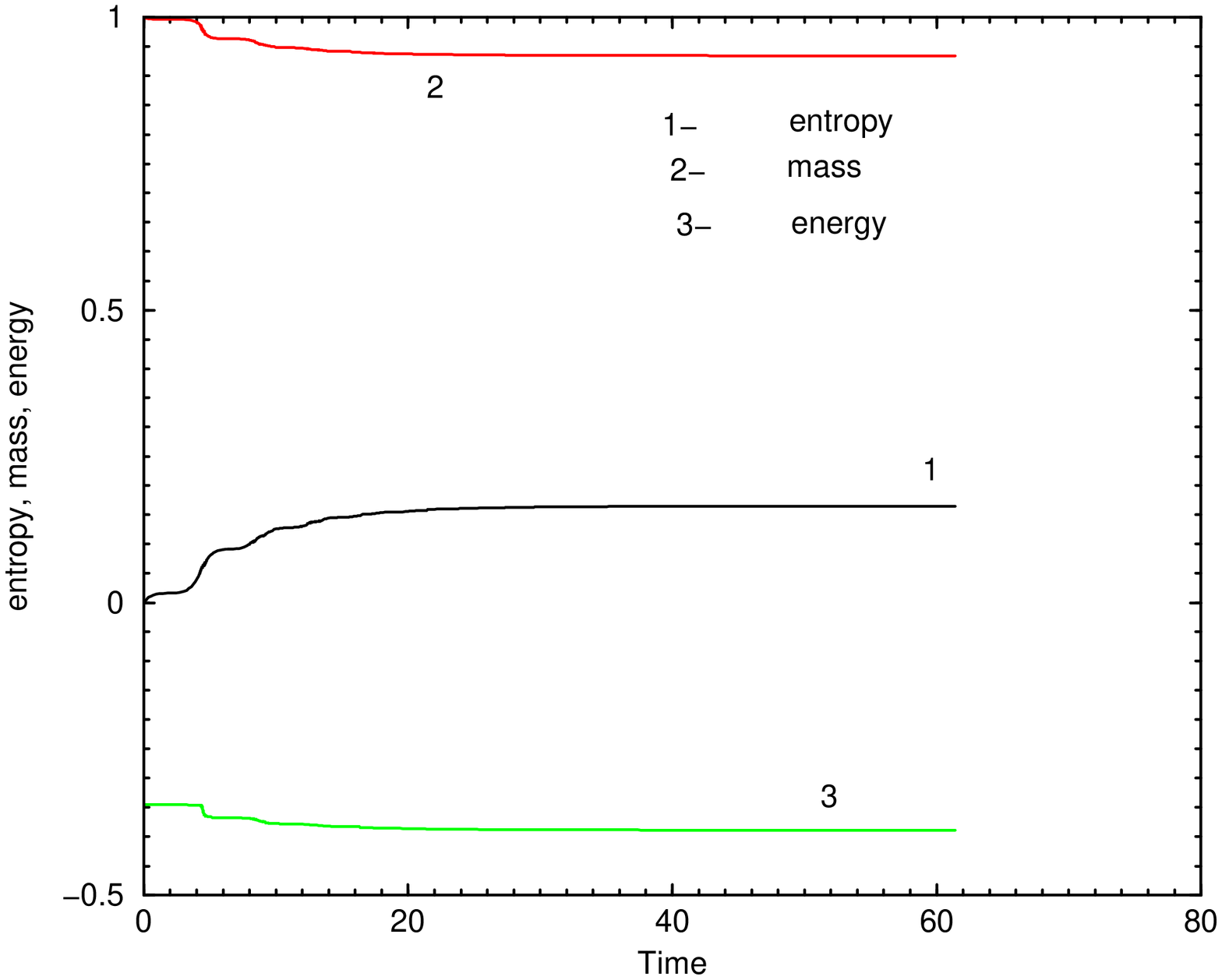,width=10cm,angle=-90}
}
\caption{Same as in Fig.2, for initial values of the variant 5, Table 1.} 
\label{fig5a}
\end{figure}

\begin{figure}
\centerline
{
\psfig{figure=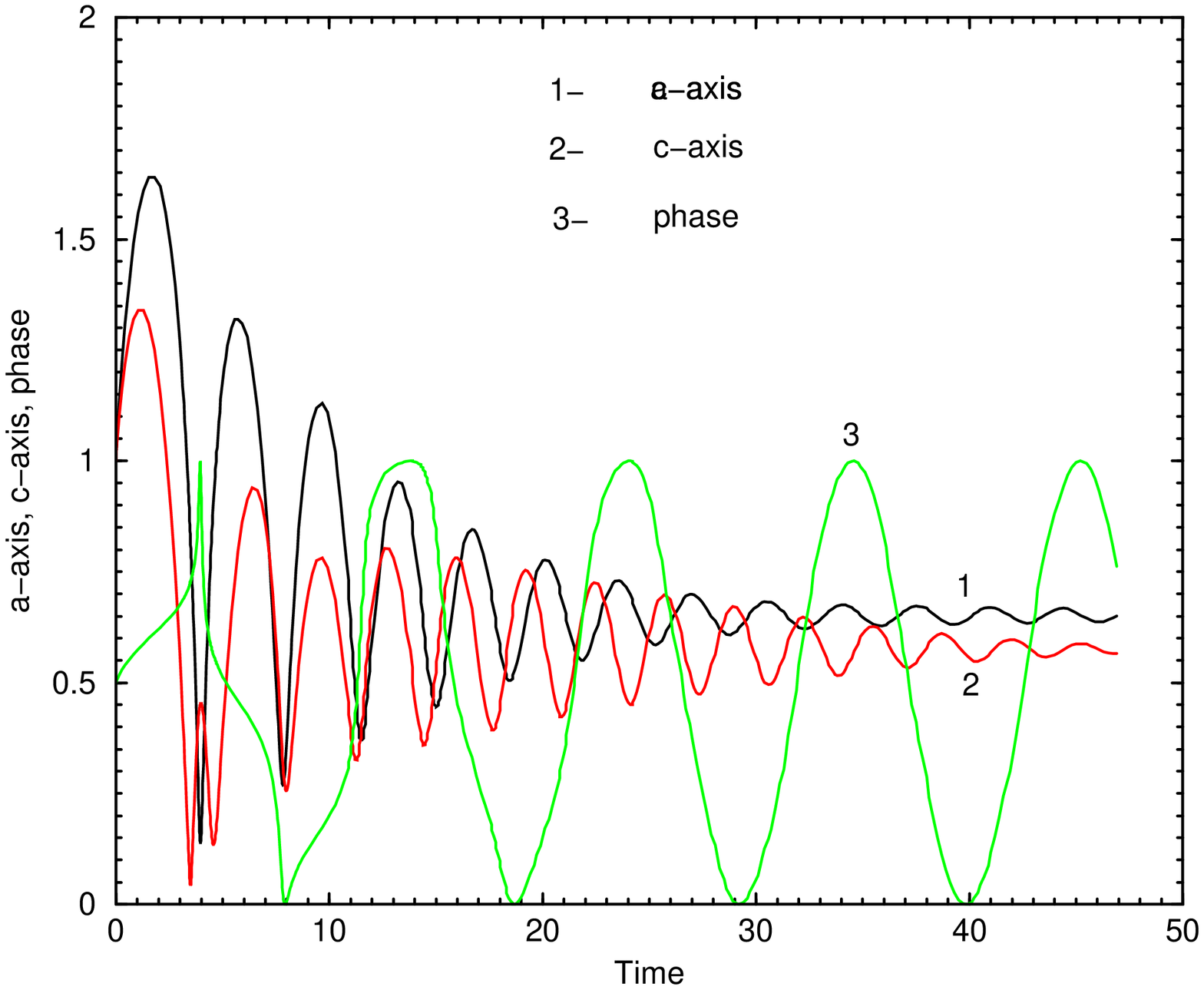,width=11cm,angle=-90}
}
\caption{Same as in Fig.1, for initial values of the variant 6, Table 1. }
\label{fig6}
\end{figure}

\begin{figure}
\centerline{
\psfig{file=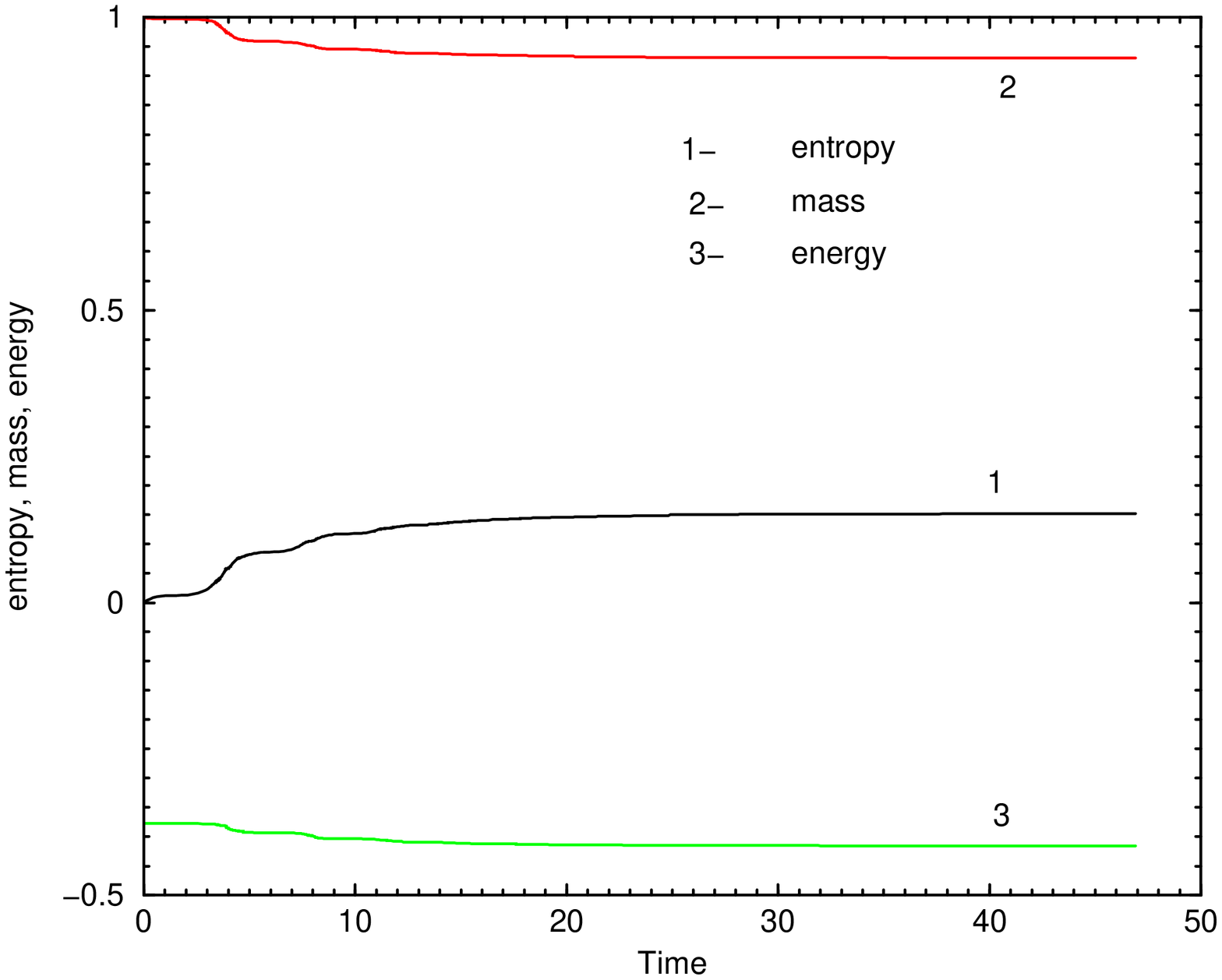,width=10cm,angle=-90}
}
\caption{Same as in Fig.2, for initial values of the variant 6, Table 1.} 
\label{fig6a}
\end{figure}

\begin{figure}
\centerline
{
\psfig{figure=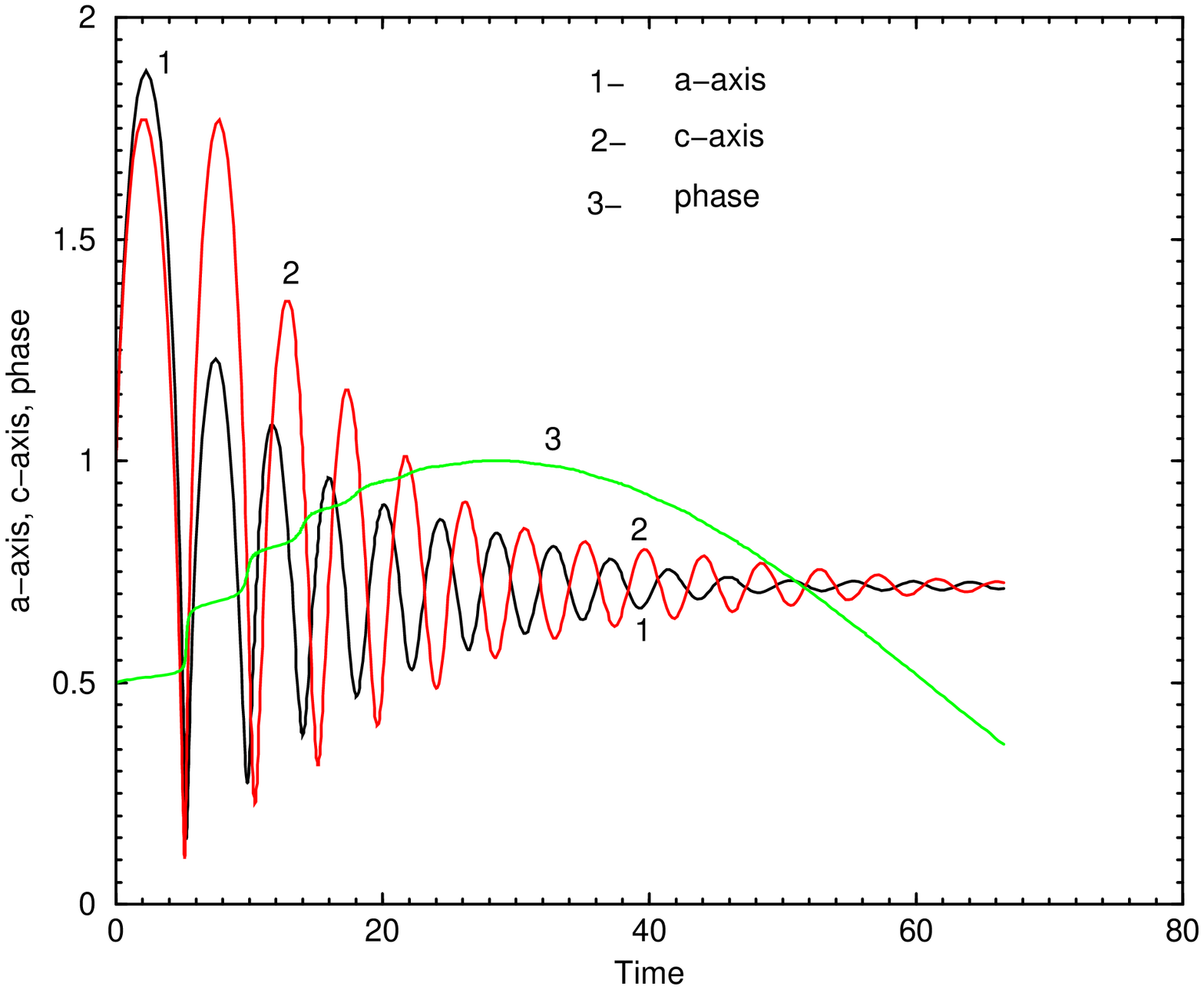,width=11cm,angle=-90}
}
\caption{Same as in Fig.1, for initial values of the variant 7, Table 1. }
\label{fig7}
\end{figure}

\begin{figure}
\centerline{
\psfig{file=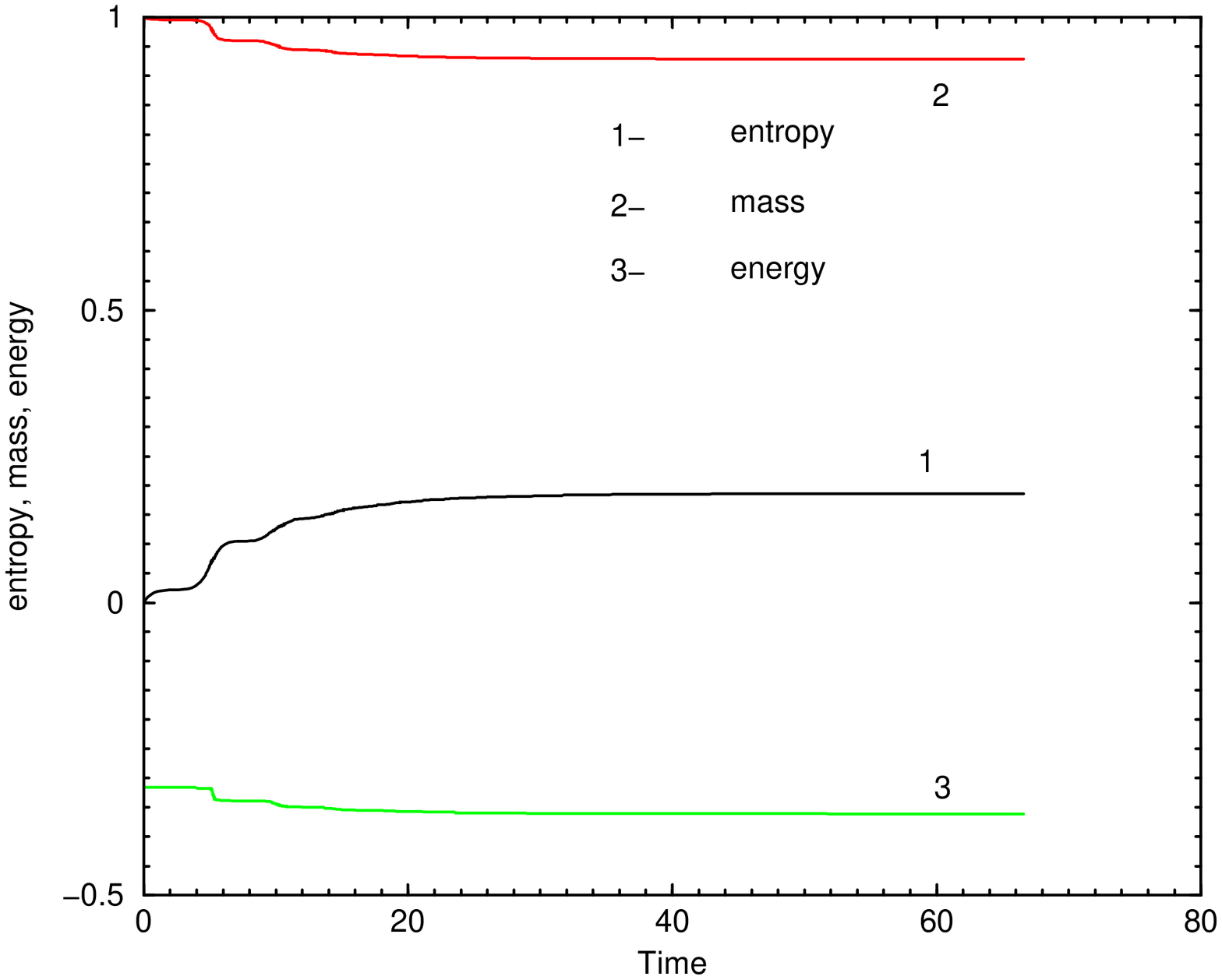,width=10cm,angle=-90}
}
\caption{Same as in Fig.2, for initial values of the variant 7, Table 1.} 
\label{fig7a}
\end{figure}

\begin{figure}
\centerline
{
\psfig{figure=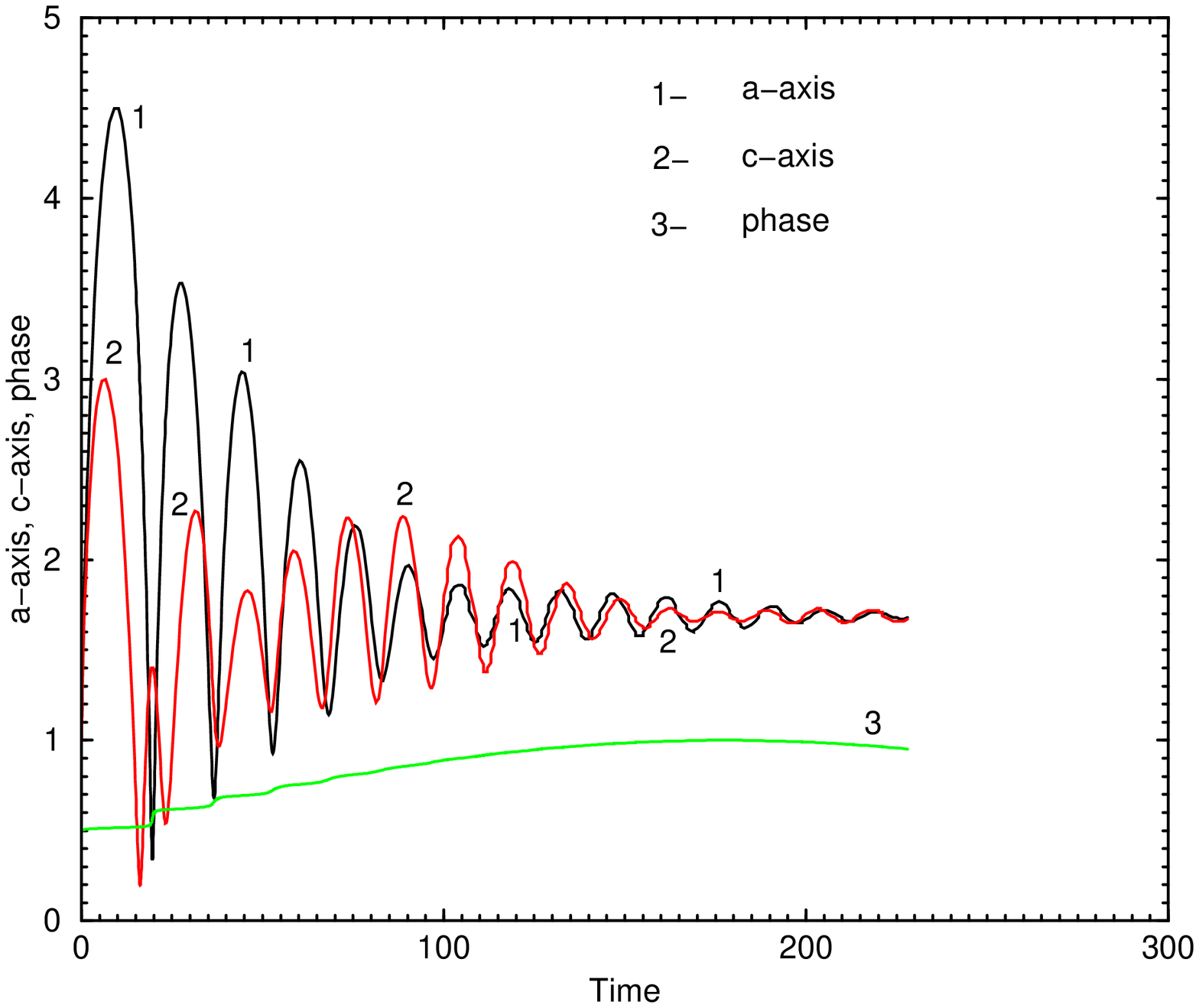,width=11cm,angle=-90}
}
\caption{Same as in Fig.1, for initial values of the variant 8, Table 1. }
\label{fig8}
\end{figure}

\begin{figure}
\centerline{
\psfig{file=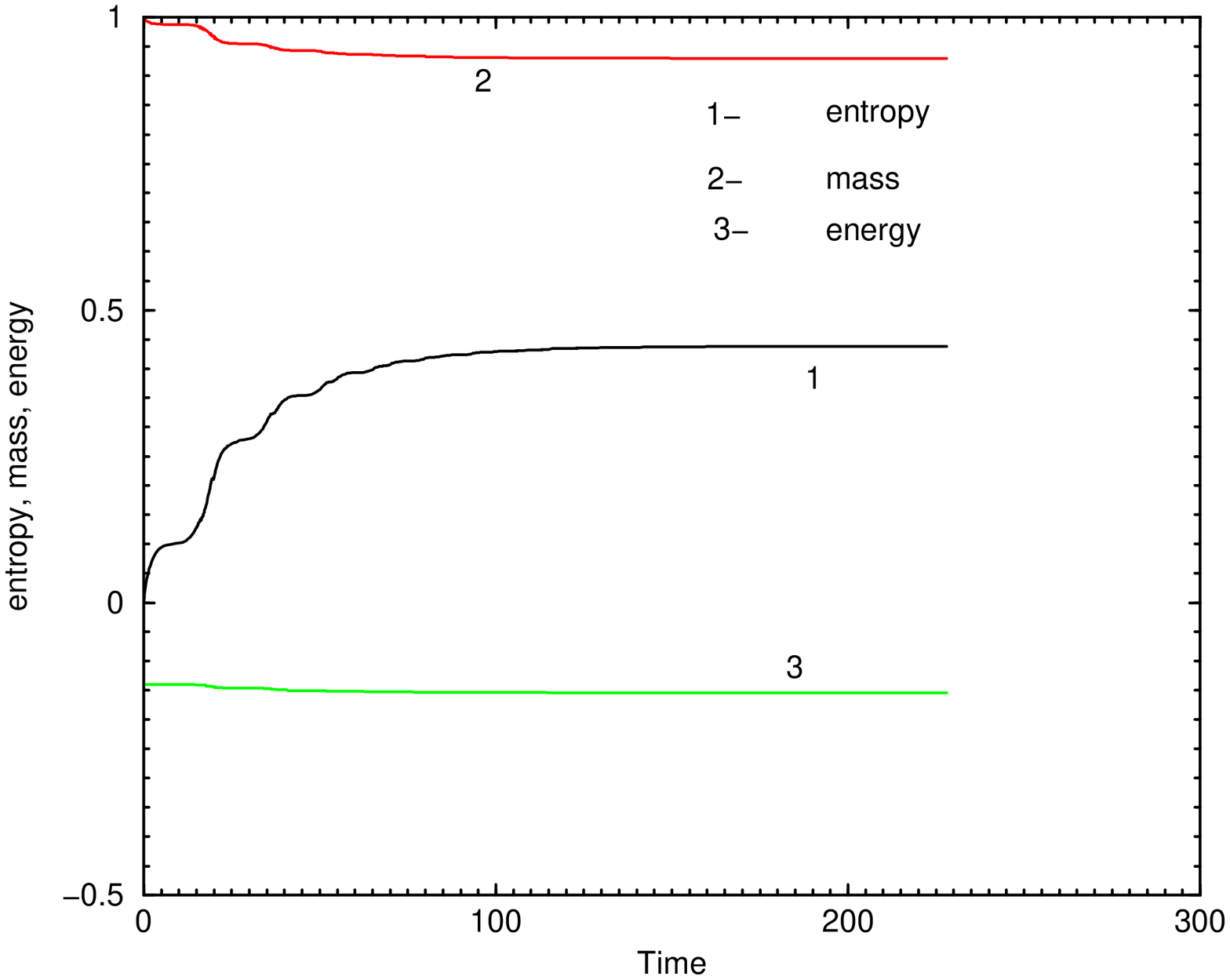,width=10cm,angle=-90}
}
\caption{Same as in Fig.2, for initial values of the variant 8, Table 1.} 
\label{figa}
\end{figure}

\medskip

{\Large{\bf Acknowledgement}}

\medskip

I grateful to Theoretical Astrophysics Center (TAC), Copenhagen, for
hospitality during the work on this paper, and to A.G. Doroshkevich and P.D. Naselsky for useful discussions.

\end{document}